\documentclass[american,twoside]{article}
\usepackage{a4wide}
\usepackage[T1]{fontenc}
\usepackage[latin1]{inputenc}
\usepackage{latexsym}
\usepackage{bbm}
\usepackage{babel}
\usepackage{times}
\usepackage{epsfig}
\usepackage{enumerate}
\usepackage{color}
\usepackage[active]{srcltx}
\usepackage[colorlinks=true]{hyperref}
\usepackage{fancyhdr}
\usepackage{mathrsfs}
\usepackage{amsmath,amsfonts,amssymb,amsthm}
\usepackage{fancyhdr}
\usepackage{graphicx}
\usepackage{bbm}
\usepackage{cancel}

\def \beq{\begin{equation}}
\def \eeq{\end{equation}}
\def \beq{\begin{equation}}
\def \eeq{\end{equation}}

\def\and {{\rm \; and \;}}

\renewcommand{\Im}{{\rm Im}\,}
\renewcommand{\Re}{{\rm Re}\,}

\newcommand{\slim}{\mathop{\mathrm{s-lim}}\limits}

\def\cS{{\cal S}}
\def\cR{{\cal R}}
\def\cH{{\cal H}}
\def\fh{{\mathfrak h}}
\def\nn{{\mathbb N}}

\def\rr{{\mathbb R}}
\def\cc{{\mathbb C}}

\def\d{{\rm d}}
\def\e{{\rm e}}
\def\i{{\rm i}}
\def\tr{{\rm tr}}
\def\sp{{\rm sp}}
\def\ds{\displaystyle}
\def\CAR{{\rm CAR}}
\def\CARv{{\rm CAR}_\vartheta}
\def\bbeta{{\boldsymbol{\beta}}}
\def\bmu{{\boldsymbol{\mu}}}
\def\bv{{\boldsymbol{v}}}
\def\b0{{\boldsymbol{0}}}
\def\st{{\langle\,\cdot\,\rangle}}
\def\jx{{\langle x\rangle}}

\newtheorem{theorem}{Theorem}[section]

\newtheorem{lemma}[theorem]{Lemma}

\theoremstyle{definition}

\numberwithin{equation}{section}

\begin{document}
\def\today{}
\title{\bf On the steady state correlation functions\\ of open interacting systems}
\author{\sc H.D. Cornean$^{1}$, V. Moldoveanu$^{2}$, C.-A. Pillet$^3$
\\ \\ \\
$^1$Department of Mathematical Sciences\\ 
Aalborg University\\
Fredrik Bajers Vej 7G, 9220 Aalborg, Denmark
\\ \\
$^2$National Institute of Materials Physics\\
P.O. Box MG-7  Bucharest-Magurele, Romania
\\ \\
$^3$
Aix-Marseille Universit\'e, CNRS UMR 7332, CPT, 13288 Marseille, France\\
Universit\'e de Toulon, CNRS UMR 7332, CPT, 83957 La Garde, France\\
FRUMAM
}
\maketitle
\thispagestyle{empty}
\begin{quote}
{\bf Abstract.} 
We address the existence of steady state Green-Keldysh correlation functions of interacting fermions
in mesoscopic systems for both the partitioning and partition-free scenarios. Under some spectral 
assumptions on the non-interacting model and for sufficiently small interaction strength, we show that 
the system evolves to a NESS which does not depend on the profile of the time-dependent 
coupling strength/bias. For the partitioned setting we also show that the steady state is independent 
of the initial state of the inner sample. Closed formulae for the NESS two-point correlation functions 
(Green-Keldysh functions), in the form of a convergent expansion, are derived. In the partitioning 
approach, we show that the $0^{\rm th}$ order term in the interaction strength of the charge current 
leads to the Landauer-B\"uttiker formula, while the $1^{\rm st}$ order correction contains the 
mean-field (Hartree-Fock) results.
\end{quote}

\maketitle
\thispagestyle{empty}

\section{Introduction and motivation}

The mathematical theory of quantum transport has attracted a lot of interest over the last decade
and substantial progress has been gradually achieved. While the development of transport theory
in condensed matter physics has been essentially geared towards computational techniques,
the fundamental question of whether a given confined system---the {\sl sample}---relaxes towards 
a stationary state when coupled to large (i.e., infinitely extended) reservoirs is much more 
delicate and requires a deeper analysis. To our knowledge, the first steps in this direction are 
due to Lebowitz and Spohn 
\cite{LS,Sp} who proved the existence of a stationary state in the van Hove (weak coupling)
limit and investigated their thermodynamic properties. Their results, based on the pioneering works 
of Davies \cite{Da1,Da2,Da3} on the weak coupling limit, hold under very general conditions.
However, due to the time rescaling inherent to this technique, they only offer a very coarse
time resolution of transport phenomena. In \cite{JP1,MMS} relaxation to a steady state 
of a $N$-level system coupled to fermionic or bosonic reservoirs has been obtained without 
rescaling, for small but finite coupling strength, and under much more stringent conditions.
The steady state obtained in these works are analytic in the coupling strength, and to
zeroth order they coincide with the weak coupling steady state of Lebowitz and Spohn.
Unfortunately, these results do not cover the particularly interesting case of confined interacting 
fermions coupled to biased non-interacting fermionic reservoirs (or leads). 

In the current paper we address the transport problem for interacting fermions in mesoscopic 
systems in two distinct situations which have been discussed in the physics literature.
\begin{enumerate}[(i)]
\item The {\sl partitioning} scenario: initially, the interacting sample is isolated from the leads and
each lead is in thermal equilibrium. At some later time $t_0$ the sample is (suddenly or adiabatically)
coupled to the leads. In this case the driving forces which induce transport are of thermodynamical 
nature: the imbalance in the leads temperatures and chemical potentials. 

\item The {\sl partition-free} scenario: the interacting sample coupled to the free leads are initially
in joint thermal equilibrium. At the later time $t_0$ a bias is imposed in each lead (again 
suddenly or adiabatically). In this case, the driving forces are of mechanical nature: the 
imbalance in the biases imposed on the leads.
\end{enumerate}

In the special case of non-interacting fermions (and the related $XY$ spin chain) in the partitioning 
scenario, conditions for relaxation to a steady state were obtained in \cite{AH,AP,AJPP2,Ne}. 
The Landauer formula was derived from the Kubo formula in \cite{CJM,CDNP}. The nonlinear 
Landauer-B\"uttiker formula was also derived in \cite{AJPP2,Ne}. See also the seminal work of 
Caroli {\it et al} \cite{CCNS} for a more physical approach. 

Non-interacting systems in the partition-free scenario were first considered by Cini \cite{Ci}. At the 
mathematical level, the existence of non-interacting steady currents and a nonlinear 
Landauer-B\"uttiker formula was also established in this setting \cite{CNZ,CGZ}. 

The physical results of Cini and Caroli {\it et al.} did not include an important ingredient: the interaction
between electrons. This last step was first achieved by Meir and Wingreen \cite{MW}. They used the
partitioning approach and the non-equilibrium Green-Keldysh functions \cite{Ke} to write down a
formula for the steady state current through an interacting region. Later on their results were 
extended to time-dependent transport \cite{JWM}. The Keldysh formalism is nowadays the standard 
tool of physicists for 
transport calculations in the presence of electron-electron interactions both for steady state and 
transient regime (see e.g. \cite{MSSL,TR}). The main reason for this is that the Keldysh-Green 
functions can be calculated from systematic many-body perturbative schemes. Nevertheless, the
Keldysh formalism for transport does not provide any arguments on the actual 
existence of the steady state, especially in the interacting case where any explicit calculation
includes approximations of the interaction self-energy. It should be mentioned here that recent 
numerical simulations using time-dependent density functional (TDDFT) methods suggest that 
systems with Hubbard-type interactions do not evolve towards a steady state \cite{KSKVG}.  
Moreover, it was also shown \cite{PFVA} that different approximation 
schemes for the interaction self-energy lead to different values of the long-time current.

Relaxation to a steady state for weakly interacting systems in the partitioning scenario 
with sudden coupling where first obtained in \cite{FMU,JOP3,MCP}. The off-resonant regime 
was investigated in \cite{CM}. The Green-Kubo formula was proven in an abstract setting along with 
the Onsager Reciprocity Relations in \cite{JOP1,JOP2} and subsequently applied to interacting
fermions \cite{JOP3}.

Our present work extends these results and treats the partitioning and partition-free 
scenarios on an equal footing. We also show that the adiabatic and sudden coupling procedures 
lead to the same results, provided some spectral condition on the non-interacting one-body 
Hamiltonian is satisfied.

We follow the scattering approach to the construction of non-equilibrium steady state advocated by 
Ruelle in \cite{Ru1,Ru2} (see also \cite{AJPP1} for a pedagogical exposition). Our analysis combines 
the Dyson expansion techniques developed in \cite{BM,FMU,BMa,JOP3} with local decay estimates
of the one-particle Hamiltonian \cite{JK}. We obtain an explicit expression for the non-equilibrium
steady state in the form of a convergent expansion in powers of the interaction strength. We 
show that this steady state does not depend on the way the coupling to the leads or the bias are
switched on. Specializing our expansion to the Green-Keldysh correlation functions, we derive a few 
basic properties of the latter and relate them to the spectral measures of a Liouvillian describing the 
dynamics of the system in the GNS representation. We also briefly discuss the Hartree-Fock 
approximation and entropy production.

The paper is organized as follows: Section \ref{sectiunea2} introduces the setting and notation. 
Section \ref{sectiunea3} contains the formulation of our main results, while Section \ref{sectiunea4}
gives their detailed proofs. Finally, in Section \ref{sectiunea6} we present our conclusions and outline 
a few open problems. 

\noindent{\bf Acknowledgments.}
HC received partial support from the Danish FNU grant {\it Mathematical Analysis of Many-Body
Quantum Systems}. VM was supported from PNII-ID-PCE Research Program (Grant No. 103/2011).
The research of CAP was partly supported by ANR (grant 09-BLAN-0098). He is grateful for the 
hospitality of the Department of Mathematical Sciences at Aalborg University, where part of this work 
was done.

\section{The model}\label{sectiunea2}
\subsection{The one-particle setup}

\begin{figure}[htbp]
\begin{center}
    \includegraphics[width=\textwidth]{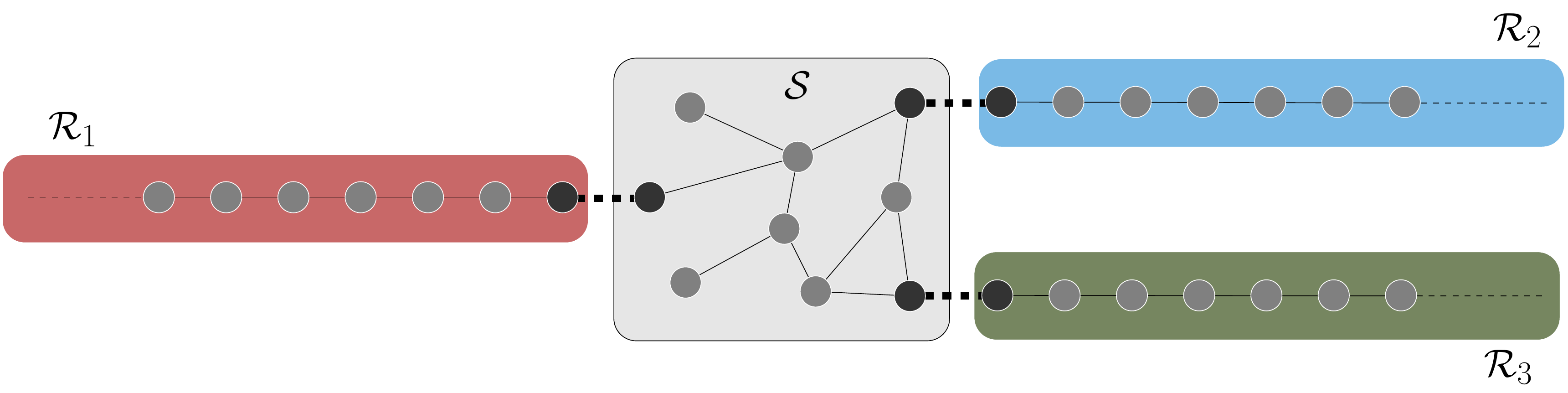}
    \caption{A finite sample $\cS$ connected to infinite reservoirs $\cR_1, \cR_2,\ldots$.}
    \label{fig:ebbm}
\end{center}
\end{figure}

We consider a Fermi gas on a discrete structure $\cS+\cR$ (e.g., an electronic system in the
tight-binding approximation). There, $\cS$ is a finite set describing a confined sample and
$\cR=\cR_1+\cdots+\cR_m$ is a collection of infinitely extended reservoirs (or leads) which feed the
sample $\cS$ (See Fig.\;\ref{fig:ebbm}). For simplicity, we will assume that these reservoirs are 
identical semi-infinite one-dimensional regular lattices. However, our approach can easily be adapted
to other geometries.

The one-particle Hilbert space of the compound system is
$$
\fh=\fh_\cS\oplus\left(\oplus_{j=1}^m\fh_j\right),
$$
where $\fh_\cS=\ell^2(\cS)$ and $\fh_j=\ell^2(\nn)$. Let $h_\cS$, a self-adjoint operator on
$\fh_\cS$, be the one-particle Hamiltonian of the isolated sample. Denote by $h_j$
the discrete Dirichlet Laplacian on $\nn$ with hopping constant $c_\cR>0$,
$$
(h_j\psi)(x)=\left\{\begin{array}{ll}
-c_\cR\,\psi(1),&\text{for } x=0,\\[4pt]
-c_\cR\left(\psi(x-1)+\psi(x+1)\right),&\text{for } x>0.
\end{array}
\right.
$$
The one-particle Hamiltonian of the reservoirs is
$$
h_\cR=\oplus_{j=1}^m h_j,
$$
and that of the decoupled system is
$$
h_{\rm D}=h_\cS\oplus h_\cR.
$$
The coupling of the sample to the reservoirs is achieved by the tunneling Hamiltonian
\begin{equation}\label{april1}
h_{\rm T}=\sum_{j=1}^m d_j\left(|\delta_{0_j}\rangle\langle\phi_j|
+|\phi_j\rangle\langle\delta_{0_j}|\right),
\end{equation}
where $\delta_{0_j}\in\fh_j$ denotes the Kronecker delta at site $0$ in $\cR_j$, 
$\phi_j\in\fh_\cS$ is a unit vector and $d_j\in\rr$ a coupling constant. The
one-particle Hamiltonian of the fully coupled system is
$$
h_\b0=h_{\rm D}+h_{\rm T}.
$$
To impose biases to the leads, the one-particle Hamiltonian of the
reservoirs and of the decoupled and coupled system will be changed to
$$
h_{\cR,\bv}=h_\cR+\left(\oplus_{j=1}^m v_j 1_j\right),\qquad
h_{{\rm D},\bv}=h_{\rm D}+\left(\oplus_{j=1}^m v_j 1_j\right),
\qquad
h_\bv=h_{{\rm D},\bv}+h_{\rm T},
$$
where $v_j\in\rr$ is the bias imposed on lead $\cR_j$, $\bv=(v_1,\ldots,v_m)$, 
and $1_j$ denotes the identity on $\fh_j$. In the following, we will
identify $1_j$ with the corresponding orthogonal projection acting in full one-particle Hilbert 
space $\fh$. The same convention applies to the identity $1_{\cS/\cR}$ on the Hilbert space
$\fh_{\cS/\cR}$.

\subsection{The many-body setup}

We shall now describe the Fermi gas associated to the one-particle model introduced previously
and extend this model by adding many-body interactions between the particles in the sample $\cS$.
In order to fix our notation and make contact with that used in the physics literature let us
recall some basic facts. We refer to \cite{BR2} for details on the algebraic framework of quantum 
statistical mechanics that we use here.

$\Gamma_-(\fh)$ denotes the fermionic Fock space over $\fh$ and
$\Gamma_-^{(n)}(\fh)=\fh^{\wedge n}$, the $n$-fold antisymmetric tensor power of $\fh$,
is the $n$-particle sector of $\Gamma_-(\fh)$.
For $f\in\fh$, let $a(f)/a^\ast(f)$ be the annihilation/creation operator on $\Gamma_-(\fh)$. 
In the following $a^\#$ stands for either $a$ or $a^\ast$. The map $f\mapsto a^\ast(f)$ is
linear while $f\mapsto a(f)$ is anti-linear, both maps being continuous, $\|a^\#(f)\|=\|f\|$.
The underlying algebraic structure is charaterized by the canonical anticommutation relations
$$
\{a(f),a^\ast(g)\}=\langle f|g\rangle,\qquad\{a(f),a(g)\}=0,
$$
and we denote by $\CAR(\fh)$ the $C^\ast$-algebra generated by $\{a^\#(f)\,|\,f\in\fh\}$, i.e., the
norm closure of the set of polynomials in the operarors $a^\#(f)$.
Note that if $\mathfrak{g}\subset\fh$ is a subspace, then we can identify $\CAR(\mathfrak{g})$
with a subalgebra of $\CAR(\fh)$.

The second quantization of a unitary operator $u$ on $\fh$ is the unitary $\Gamma(u)$ on
$\Gamma_-(\fh)$ acting as $u\otimes u\otimes\cdots\otimes u$ on $\Gamma_-^{(n)}(\fh)$.
The second quantization of a self-adjoint operator $q$ on $\fh$ is the self-adjoint generator
$\d\Gamma(q)$ of the strongly continuous unitary group $\Gamma(\e^{\i tq})$, i.e.,
$\Gamma(\e^{\i tq})=\e^{\i t\d\Gamma(q)}$. If $\{f_\iota\}_{\iota\in I}$ is an 
orthonormal basis of $\fh$ and $q$ a bounded self-adjoint operator, then
$$
\d\Gamma(q)=\sum_{\iota,\iota'\in I}\langle f_\iota|q f_{\iota'}\rangle 
a^\ast(f_\iota)a(f_{\iota'}),
$$
holds on $\Gamma_-(\fh)$. In particular, if $q$ is trace class, then $\d\Gamma(q)\in\CAR(\fh)$.

A unitary operator $u$ on $\fh$ induces a Bogoliubov automorphism of $\CAR(\fh)$
$$
A\mapsto \gamma_u(A)=\Gamma(u)A\Gamma(u)^\ast,
$$
such that $\gamma_u(a^\#(f))=a^\#(uf)$. If $t\mapsto u_t$ is a strongly continuous family
of unitary operators on $\fh$, then $t\mapsto\gamma_{u_t}$ is a strongly continuous family of
Bogoliubov automorphisms of $\CAR(\fh)$. In particular, if $u_t=\e^{\i tk}$ for some self-adjoint 
operator $k$ on $\fh$, we call $\gamma_{u_t}$ the quasi-free dynamics generated by $k$. 

The quasi-free dynamics generated by the identity $1$ is the gauge group of $\CAR(\fh)$
and  $N=\d\Gamma(1)$ is the number operator on $\Gamma_-(\fh)$,
$$
\vartheta^t(a^\#(f))=\e^{\i tN}a^\#(f)\e^{-\i tN}=a^\#(\e^{\i t}f)=\left\{
\begin{array}{rl}
\e^{-\i t}a(f)&\text{for }a^\#=a;\\[4pt]
\e^{\i t}a^\ast(f)&\text{for }a^\#=a^\ast.
\end{array}
\right.
$$
The algebra of observables of the Fermi gas is the gauge-invariant subalgebra of $\CAR(\fh)$,
$$
\CARv(\fh)=\{A\in\CAR(\fh)\,|\,\vartheta^t(A)=A\text{ for all }t\in\rr\}.
$$
It is the $C^\ast$-algebra generated by the set of all monomials in the $a^\#$ containing an 
equal number of $a$ and $a^\ast$ factors. Note that the map
$$
\mathfrak{p}_\vartheta(A)=\int_0^{2\pi}\vartheta^t(A)\frac{\d t}{2\pi},
$$
is a norm 1 projection onto $\CARv(\fh)$. Thus, as a Banach space, $\CAR(\fh)$ is the
direct sum of $\CARv(\fh)$ and its complement, the range of $(\rm{id}-\mathfrak{p}_\vartheta)$.

\subsubsection{Interacting dynamics}

The quasi-free dynamics generated by $h_\bv$  describes the sample coupled to the leads and
$H_\bv=\d\Gamma(h_\bv)$ is the corresponding many-body Hamiltonian
$$
\tau_{H_\bv}^t(a^\#(f))=\e^{\i tH_\bv}a^\#(f)\e^{-\i tH_\bv}=a^\#(\e^{\i th_\bv}f).
$$
The group $\tau_{H_\bv}$ commutes with the gauge group $\vartheta$ so that it leaves 
$\CARv(\fh)$ invariant. In the following, we shall consistently denote one-particle
operators with lower-case letters and capitalize the corresponding second quantized operator,
e.g., $H_\cS=\d\Gamma(h_\cS)$, $H_\cR=\d\Gamma(h_\cR)$, etc. We shall also denote
the corresponding groups of automorphism by $\tau_{H_\cS}$, $\tau_{H_\cR}$, etc.

We allow for interactions between particles in the sample $\cS$. However, particles
in the leads remain free. The interaction energy within the sample is described by
$$
W=\sum_{k\ge2}\frac1{k!}
\sum_{x_1,\ldots,x_k\in\cS}\Phi^{(k)}(x_1,\ldots,x_k)n_{x_1}\cdots n_{x_k},
$$
where $n_x=a^\ast(\delta_x)a(\delta_x)$ and the $k$-body interaction $\Phi^{(k)}$ is a completely
symmetric real valued function on $\cS^k$ which vanishes whenever two of its arguments coincide.
Note that $W$ is a self-adjoint element of $\CARv(\fh)$. For normalization purposes, we assume that 
$|\Phi^{(k)}(x_1,\ldots,x_k)|\le1$.
A typical example is provided by the second quantization of a pair potential
$w(x,y)=w(y,x)$ describing the interaction energy between two particles at sites $x,y\in\cS$.
The corresponding many-body operator is
\begin{equation}
W=\frac1{2}\sum_{x,y\in\cS}w(x,y)n_xn_y.
\label{Vint}
\end{equation}
For any self-adjoint $W\in\CARv(\fh)$ and  any value of the interaction strength $\xi\in\rr$ the operator
$$
K_\bv=H_\bv+\xi W,
$$
is self-adjoint on the domain of $H_\bv$.
Moreover $\tau_{K_\bv}^t(A)=\e^{\i tK_\bv}A\e^{-\i tK_\bv}$ defines a strongly continuous group
of $\ast$-automorphisms of $\CAR(\fh)$ leaving invariant the subalgebra $\CARv(\fh)$.
This group describes the full dynamics of the Fermi gas, including interactions. It has the following
norm convergent Dyson expansion
$$
\tau_{K_\bv}^t(A)=\tau_{H_\bv}^t(A)
+\sum_{n=1}^\infty(\i\xi)^n\int\limits_{0\le s_1\le\cdots\le s_n\le t}
[\tau_{H_\bv}^{s_1}(W),[\tau_{H_\bv}^{s_2}(W),[\cdots,
[\tau_{H_\bv}^{s_n}(W),\tau_{H_\bv}^t(A)]\cdots]]]
\d s_1\cdots\d s_n.
$$

\subsubsection{States of the Fermi gas}
\label{FermiState}

A state on $\CAR(\fh)$ is a linear functional
$$
\begin{array}{ccc}
\CAR(\fh)&\to&\cc\\
A&\mapsto&\langle A\rangle,
\end{array}
$$ 
such that $\langle A^\ast A\rangle\ge0$ for all $A$ and $\langle I\rangle=1$. A state is gauge-invariant 
if $\langle\vartheta^t(A)\rangle=\langle A\rangle$ for all $t\in\rr$. Note that if $\st$ is 
a state on $\CAR(\fh)$ then its restriction to $\CARv(\fh)$ defines a state on this subalgebra. We shall 
use the same notation for this restriction. Reciprocally, if $\st$ is a state on 
$\CARv(\fh)$ then $\langle\mathfrak{p}_\vartheta(\,\cdot\,)\rangle$ is a gauge-invariant state on 
$\CAR(\fh)$. 

A state $\st$ on $\CAR(\fh)$ induces a GNS representation $(\cH,\pi,\Omega)$
where $\cH$ is a Hilbert space, $\pi$ is a $\ast$-morphism from $\CAR(\fh)$ to the bounded
linear operators on $\cH$ and $\Omega\in\cH$ is a unit vector such that $\pi(\CAR(\fh))\Omega$ is 
dense in $\cH$ and $\langle A\rangle=(\Omega|\pi(A)\Omega)$ for all $A\in\CAR(\fh)$. Let
$\rho$ be a density matrix on $\cH$ (a non-negative, trace class operator with $\tr(\rho)=1$).
The map $A\mapsto\tr(\rho\pi(A))$ defines a state on $\CAR(\fh)$. Such a state is said to be normal
w.r.t.\;$\st$. From the thermodynamical point of view $\st$-normal states are close to $\st$ and 
describe local perturbations of this state.

Given a self-adjoint operator $\varrho$ on $\fh$ satisfying $0\le\varrho\le I$, the formula
\beq
\langle a^\ast(f_1)\cdots a^\ast(f_k)a(g_l)\cdots a(g_1)\rangle_\varrho
=\delta_{kl}\det\{\langle g_j|\varrho f_i\rangle\},
\label{GIQF}
\eeq
defines a unique gauge-invariant state on $\CAR(\fh)$. This state is called the quasi-free state of 
density $\varrho$. It is uniquely determined by the two point function
 $\langle a^\ast(f)a(g)\rangle_\varrho=\langle g|\varrho f\rangle$. An alternative characterization
of quasi-free states on $\CAR(\fh)$ is the usual fermionic Wick theorem
\beq
\langle\varphi(f_1)\cdots\varphi(f_k)\rangle_\varrho=\left\{
\begin{array}{ll}
0,&\text{if $k$ is odd;}\\[8pt]
\ds\sum_{\pi\in\mathcal{P}_k}\varepsilon(\pi)
\prod_{j=1}^{k/2} \langle\varphi(f_{\pi(2j-1)})\varphi(f_{\pi(2j)})\rangle_\varrho,&\text{if $k$ is even};
\end{array}
\right.
\label{wickform}
\eeq
where $\varphi(f)=2^{-1/2}(a^\ast(f)+a(f))$ is the field operator, $\mathcal{P}_k$ denotes the
set of pairings of $k$ objects, i.e., permutations satisfying $\pi(2j-1)<\min(\pi(2j),\pi(2j+1))$
for $j=1,\ldots,k/2$, and $\epsilon(\pi)$ is the signature of the permutation $\pi$.

Given a strongly continuous group $\tau$ of $\ast$-automorphisms of $\CAR(\fh)$ commuting with the 
gauge group $\vartheta$, a state $\st$ is a thermal equilibrium state at inverse
temperature $\beta$ and chemical potential $\mu$ if it satisfies the $(\beta,\mu)$-KMS condition 
w.r.t.\;$\tau$, i.e., if for any $A,B\in\CAR(\fh)$ the function
$$
F_{A,B}(t)=\langle A\tau^t\circ\vartheta^{-\mu t}(B)\rangle,
$$
has an analytic continuation to the strip $\{0<\Im t<\beta\}$ with a bounded continuous extension
to the closure of this strip satisfying
$$
F_{A,B}(t+\i\beta)=\langle\tau^t\circ\vartheta^{-\mu t}(B)A\rangle.
$$
We shall say that such a state is a $(\beta,\mu)$-KMS state for $\tau$. 


Let $k$ be a self-adjoint operator on $\fh$ and $K=\d\Gamma(k)$. For any $\beta,\mu\in\rr$, the 
quasi-free dynamics $\tau_K$ generated by $k$ has a unique $(\beta,\mu)$-KMS state: the quasi-free 
state with density $\varrho_k^{\beta,\mu}=(I+\e^{\beta(k-\mu)})^{-1}$ which we shall denote by 
$\st_K^{\beta,\mu}$. If $Q\in\CARv(\fh)$ is self-adjoint then 
$K_Q=K+Q$ generates a strongly continuous group $\tau_{K_Q}$ of $\ast$-automorphisms of
$\CAR(\fh)$ leaving the subalgebra $\CAR_v(\fh)$ invariant. It follows from Araki's perturbation 
theory that $\tau_{K_Q}$ also has a unique $(\beta,\mu)$-KMS state denoted
$\st_{K_Q}^{\beta,\mu}$. Moreover, this state is normal 
w.r.t.\;$\st_K^{\beta,\mu}$.
In particular, the coupled non-interacting dynamics 
$\tau_{H_\b0}$ and the coupled interacting dynamics $\tau_{K_\b0}$ have unique $(\beta,\mu)$-KMS 
states $\st_{H_\b0}^{\beta,\mu}$ and  $\st_{K_\b0}^{\beta,\mu}$
which are mutually normal.

\noindent{\bf Remark 1.} It is well known that for any $\beta>0$ and $\mu\in\rr$ the KMS states 
$\st_{H_\b0}^{\beta,\mu}$ and $\st_{K_\b0}^{\beta,\mu}$ are thermodynamic limits of the familiar 
grand canonical Gibbs states associated to the restrictions of the Hamiltonian $H_\b0$ and $K_\b0$ 
to finitely extended reservoirs with appropriate boundary conditions. See \cite{BR2} for details.

\noindent{\bf Remark 2.} If the Hamiltonian $h_\cS$ and the coupling functions $\phi_j$ are such that
$\langle\delta_x|h_\cS\delta_y\rangle$ and $\langle\delta_x|\phi_j\rangle$ are real
for all $x,y\in\cS$ then the $C^\ast$-dynamics $\tau_{K_\bv}$ is time reversal invariant.
More precisely, let $\Theta$ be the anti-linear involutive $\ast$-automorphism of $\CAR(\fh)$ defined by 
$\Theta(a^\#(f))=a^\#(\overline{f})$, where $\overline{\vphantom{f}\,\cdot\,}$ denotes the natural 
complex conjugation on the one-particle Hilbert space $\fh=\ell^2(\cS)\oplus(\oplus_{j=1}^m\ell^2(\nn))$.
Then one has
\beq
\tau_{K_\bv}^t\circ\Theta=\Theta\circ\tau_{K_\bv}^{-t},
\label{TRIDyn}
\eeq
for all $t\in\rr$. The same remark holds for the non-interacting dynamics $\tau_{H_\bv}$ and
for the decoupled dynamics $\tau_{{\rm D},\bv}$. Moreover, the KMS-state 
$\st_{K_\b0}^{\beta,\mu}$ is time reversal invariant, i.e.,
$$
\langle\Theta(A)\rangle_{K_\b0}^{\beta,\mu}=\langle A^\ast\rangle_{K_\b0}^{\beta,\mu},
$$
holds for all $A\in\CAR(\fh)$. In particular $\langle A\rangle_{K_\b0}^{\beta,\mu}=0$ for
any self-adjoint $A\in\CAR(\fh)$ such that $\Theta(A)=-A$.

\subsubsection{Current observables}

Physical quantities of special interest are the charge and energy currents through the sample $\cS$.
To associate observables (i.e., elements of $\CARv(\fh)$)
to these quantities, note that the total charge inside lead $\cR_j$ 
is described by $N_j=\d\Gamma(1_j)$. Even though this operator is not an
observable (and has infinite expectation in a typical state like 
$\st_{H_\b0}^{\beta,\mu}$), its time derivative
\begin{align}
J_j=
-\left.\frac{\d\ }{\d t}\,\e^{\i tK_\bv}N_j\e^{-\i tK_\bv}\right|_{t=0}&=-\i[K_\bv,N_j]
=-\i[\d\Gamma(h_{{\rm D},\bv}+h_{\rm T})+\xi W,\d\Gamma(1_j)]\nonumber\\
&=\d\Gamma(\i[1_j,h_{\rm T}])
=\i d_j\left(a^\ast(\delta_{0_j})a(\phi_j)-a^\ast(\phi_j)a(\delta_{0_j})\right),
\label{Jdef}
\end{align}
belongs to $\CARv(\fh)$ and hence
$$
\e^{\i tK_\bv}N_j\e^{-\i tK_\bv}-N_j=-\int_0^t\tau_{K_\bv}^s(J_j)\,\d s,
$$
is an observable describing the net charge transported into lead $\cR_j$ during the period
$[0,t]$. We shall consequently consider $J_j$ as the charge current entering the sample
from lead $\cR_j$. Gauge invariance implies that the total charge inside the sample,
which is described by the observable $N_\cS=\d\Gamma(1_\cS)\in\CARv(\fh)$, satisfies
$$
\left.\frac{\d\ }{\d t}\tau_{K_\bv}^t(N_\cS)\right|_{t=0}=\i[K_\bv,N_\cS]
=\i\left[K_\bv,N-\sum_{j=1}^mN_j\right]=\sum_{j=1}^m J_j.
$$

In a similar way we define the energy currents
\begin{align}
E_j=
-\left.\frac{\d\ }{\d t}\,\e^{\i tK_\bv}H_j\e^{-\i tK_\bv}\right|_{t=0}&=-\i[K_\bv,H_j]
=-\i[\d\Gamma(h_{{\rm D},\bv}+h_{\rm T})+\xi W,\d\Gamma(h_j)]\nonumber\\
&=\d\Gamma(\i[h_j,h_{\rm T}])
=-\i c_\cR d_j\left(a^\ast(\delta_{1_j})a(\phi_j)-a^\ast(\phi_j)a(\delta_{1_j})\right),
\label{Edef}
\end{align}
and derive the conservation law
$$
\left.\frac{\d\ }{\d t}\tau_{K_\bv}^t(H_\cS+\xi W+H_{\rm T})\right|_{t=0}=\i[K_\bv,H_\cS+\xi W+H_{\rm T}]
=\i\left[K_\bv,K_\bv-\sum_{j=1}^m(H_j+v_jN_j)\right]=\sum_{j=1}^m E_j+v_jJ_j.
$$
It follows that for any $\tau_{K_\bv}$-invariant state $\st$ one has the
sum rules
\beq
\sum_{j=1}^m\langle J_j\rangle=0,\qquad
\sum_{j=1}^m\langle E_j+v_jJ_j\rangle=0,
\label{SumRules}
\eeq
which express charge and energy conservation. Note that, despite their definition, the current
observables do not depend on the bias $\bv$.

\noindent{\bf Remark.} Charge and energy transport within the system can also be characterized 
by the so called counting statistics (see \cite{LL, ABGK, LLL} and the 
comprehensive review \cite{EHM}). We shall not consider this option
here and refer the reader to \cite{DRM,JOPP,JOPS} for discussions and 
comparisons of the two approaches and to \cite{FNBSJ,FNBJ} for a glance on the 
problem of full counting statistics in interacting non-markovian systems.

\subsection{Non-equilibrium steady states}

Two physically distinct ways of driving the combined system $\cS+\cR$ out of equilibrium have been
used and discussed in the literature. In the first case, the {\sl partitioning scenario,} one does not 
impose any bias in the reservoirs. The latter remain decoupled from the sample at early times $t<t_0$.
During this period each reservoir is in thermal equilibrium, but the intensive thermodynamic parameters 
(temperatures and chemical potentials) of these reservoirs are distinct so that they are not in mutual 
equilibrium. At time $t=t_0$ one switches on the couplings to the sample and let the system evolve 
under the full unbiased dynamics $\tau_{K_\b0}$. In the second case, the {\sl partition-free scenario,} the combined 
system $\cS+\cR$ remains coupled at all times. For $t<t_0$ it is in a thermal equilibrium state 
associated to the unbiased dynamics $\tau_{K_\b0}$. At time $t=t_0$ a bias $\bv\not=\b0$ is applied
to the leads and the system then evolves according to the biased dynamics $\tau_{K_\bv}$. In both 
cases, it is expected that as $t_0\to-\infty$, the system will reach a steady state at time $t=0$.

We shall adopt a unified approach which allows us to deal with these two scenarios on an equal
footing and to consider mixed situations where both thermodynamical and mechanical forcing
act on the sample.

We say that the gauge-invariant state $\st$ on $\CAR(\fh)$ is 
almost-$(\bbeta,\bmu)$-KMS with
$\bbeta=(\beta_1,\ldots,\beta_m)\in\rr_+^m$ and $\bmu=(\mu_1,\ldots,\mu_m)\in\rr^m$ if it is
normal w.r.t.\;the quasi-free state $\st_\cR^{\bbeta,\bmu}$ on  $\CAR(\fh)$
with density
\beq
\varrho^{\bbeta,\bmu}_\cR=\left (\frac12 1_\cS\right ) \oplus\left(\bigoplus_{j=1}^m
\frac1{1+\e^{\beta_j(h_j-\mu_j)}}\right).
\label{rhoRbetamu}
\eeq
The restriction of $\st_\cR^{\bbeta,\bmu}$ to $\CAR(\fh_{\cR_j})$ is the unique
$(\beta_j,\mu_j)$-KMS state for $\tau_{H_j}$. Its restriction to $\CAR(\fh_{\cS})$ is the unique
tracial state on this finite dimensional algebra. We also remark that if 
$\bbeta=(\beta,\ldots,\beta)$ and $\bmu=(\mu,\ldots,\mu)$ then 
$\st_\cR^{\bbeta,\bmu}$ is the $(\beta,\mu)$-KMS state for $\tau_{H_\cR}$.

An almost-$(\bbeta,\bmu)$-KMS state describes the situation where each reservoir $\cR_j$
is near thermal equilibrium at inverse temperature $\beta_j$ and chemical potential
$\mu_j$. There is however no restriction on the state of the sample $\cS$ which can be
an arbitrary gauge-invariant state on $\CAR(\fh_\cS)$. In particular, an almost-$(\bbeta,\bmu)$-KMS 
state needs not be quasi-free or a product state.

We say that the gauge-invariant state $\st^{\bbeta,\bmu,\bv}_+$ on $\CARv(\fh)$ is 
{\sl the} $(\bbeta,\bmu,\bv)$-NESS of the system $\cS+\cR$ if
$$
\langle A\rangle^{\bbeta,\bmu,\bv}_+=\lim_{t_0\to-\infty}\langle\tau_{K_\bv}^{-t_0}(A)\rangle,
$$
holds for {\sl any} almost-$(\bbeta,\bmu)$-KMS state $\st$ and any 
$A\in\CARv(\fh)$. Since
$$
\langle\tau_{K_\bv}^{t}(A)\rangle^{\bbeta,\bmu,\bv}_+
=\lim_{t_0\to-\infty}\langle\tau_{K_\bv}^{t-t_0}(A)\rangle=
\langle A\rangle^{\bbeta,\bmu,\bv}_+,
$$
the $(\bbeta,\bmu,\bv)$-NESS, if it exists, is invariant under the full dynamics $\tau_{K_\bv}$. 
By definition, it is independent of the initial state of the system $\cS$. 

In the two following sections we explain how the partitioning and partition-free scenario
fits into this general framework. We also introduce time-dependent protocols to study
the effect of an adiabatic switching of the tunneling Hamiltonian $H_{\rm T}$ or of the
bias $\bv$.

\subsubsection{The partitioning scenario}

In this scenario, there is no bias in the leads, i.e., $\bv=\b0$ at any time. The initial
state is an almost-$(\bbeta,\bmu)$-KMS product state
$$
\langle A_\cS A_1\cdots A_m\rangle
=\langle A_\cS\rangle_\cS \langle A_1\rangle_{H_{\cR_1}}^{\beta_1,\mu_1}
\cdots\langle A_m\rangle_{H_{\cR_m}}^{\beta_m,\mu_m},
$$
for $A_\cS\in\CARv(\fh_\cS)$ and $A_j\in\CARv(\fh_j)$,
where $\st_\cS$ is an arbitrary gauge-invariant state on $\CAR(\fh_\cS)$ and
$\st_{H_{\cR_j}}^{\beta_j,\mu_j}$ is the $(\beta_j,\mu_j)$-KMS state on
$\CAR(\fh_j)$ for $\tau_{H_{\cR_j}}$.

We shall also discuss the effect of an adiabatic switching of the coupling between $\cS$ and
$\cR$. To this end, we replace the tunneling Hamiltonian $H_{\rm T}$ with the time dependent one 
$\chi(t/t_0)H_{\rm T}$ where $\chi:\rr\to[0,1]$ is a continuous function such that $\chi(t)=1$ for $t\le0$
and $\chi(t)=0$ for $t\ge1$. Thus, for $t_0<0$, the time dependent Hamiltonian 
$$
K_\b0(t/t_0)=H_\cS+H_\cR+\chi(t/t_0)H_{\rm T}+\xi W,
$$
is self-adjoint on the domain of $H_\cR$ and describes the switching of the coupling $H_{\rm T}$ 
during the time period $[t_0,0]$. It generates a strongly continuous two parameter family of unitary 
operators $U_{\b0,t_0}(t,s)$ on $\Gamma_-(\fh)$ satisfying
$$
\i\partial_tU_{\b0,t_0}(t,s)\Psi=K_\b0(t/t_0)U_{\b0,t_0}(t,s)\Psi,\quad U_{\b0,t_0}(s,s)=I,
$$
for $\Psi$ in the domain of $H_\cR$. One easily shows that the formula
$$
\alpha_{\b0,t_0}^{s,t}(A)=U_{\b0,t_0}(t,s)^\ast AU_{\b0,t_0}(t,s),
$$
defines a strongly continuous two parameter family of $\ast$-automorphisms of $\CAR(\fh)$
leaving $\CARv(\fh)$ invariant. $\alpha_{\b0,t_0}^{s,t}$ describes the non-autonomous evolution of 
the system from time $s$ to time $t$ under adiabatic coupling. It satisfies the composition rule
$$
\alpha_{\b0,t_0}^{s,u}\circ\alpha_{\b0,t_0}^{u,t}=\alpha_{\b0,t_0}^{s,t},
$$
for any $s,u,t\in\rr$, and in particular $(\alpha_{\b0,t_0}^{s,t})^{-1}=\alpha_{\b0,t_0}^{t,s}$.

\subsubsection{The partition-free scenario}

In this case the bias $\bv$ is non-zero and the initial state is a thermal equilibrium state for
the unbiased full dynamics, i.e., $\st_{K_\b0}^{\beta,\mu}$ for
some $\beta>0$ and $\mu\in\rr$. Note that since $K_\b0=H_\cR+Q$ with
$Q=H_\cS+\xi W+H_{\rm T}\in\CARv(\fh)$, this state is almost-$(\bbeta,\bmu)$-KMS with
$\bbeta=(\beta,\ldots,\beta)$ and $\bmu=(\mu,\ldots,\mu)$.

We shall also consider the adiabatic switching of the bias via the time dependent Hamiltonian
$$
K_{\bv}(t/t_0)=H_\cS+H_\cR+H_{\rm T}+\chi(t/t_0)V_\cR+\xi W,
$$
where
$$
V_\cR=\d\Gamma(v_\cR)=\sum_{j=1}^mv_j \d\Gamma(1_j).
$$
We denote by $U_{\bv,t_0}(t,s)$ the corresponding family of unitary propagators on the Fock space
$\Gamma_-(\fh)$, and define
$$
\alpha_{\bv,t_0}^{s,t}(A)=U_{\bv,t_0}(t,s)^\ast AU_{\bv,t_0}(t,s).
$$

\subsubsection{NESS Green-Keldysh correlation functions}
\label{Sct-Green}

Let $\st$ be the state of the system at time $t_0$. The so called lesser, greater, retarded and 
advanced Green-Keldysh correlation functions are defined as
\begin{align*}
G^<(t,s;x,y)&=+\i\langle\tau_{K_\bv}^{s-t_0}(a_y^\ast)\tau_{K_\bv}^{t-t_0}(a_x)\rangle,\\
G^>(t,s;x,y)&=-\i\langle\tau_{K_\bv}^{t-t_0}(a_x)\tau_{K_\bv}^{s-t_0}(a^\ast_y)\rangle,\\
G^{\rm r}(t,s;x,y)&=+\i\theta(s-t)\langle\{\tau_{K_\bv}^{s-t_0}(a_y^\ast),\tau_{K_\bv}^{t-t_0}(a_x)\}\rangle,\\
G^{\rm a}(t,s;x,y)&=-\i\theta(t-s)\langle\{\tau_{K_\bv}^{s-t_0}(a_y^\ast),\tau_{K_\bv}^{t-t_0}(a_x)\}\rangle,
\end{align*}
where we have set $a_x=a(\delta_x)$ for $x\in\cS\cup\cR$ and $\theta$ denotes the Heaviside step 
function.

A number of physically interesting quantities can be expressed in terms of these Green's functions, 
e.g., the charge density
$$
{\frak n}(x,t)=\langle\tau_{K_\bv}^{t-t_0}(a^\ast_xa_x)\rangle=\Im G^<(t,t;x,x),
$$
or the electric current out of lead $\cR_j$,
$$
{\frak j}_j(t)=\langle\tau_{K_\bv}^{t-t_0}(J_j)\rangle=-2d_j\Re\sum_{x\in\cS}G^<(t,t;0_j,x)\phi_j(x).
$$

Assuming existence of the $(\bbeta,\bmu,\bv)$-NESS, the limiting Green's functions
$\lim_{t_0\to-\infty}G^{\square}(t,s;x,y)$ only depend on the time difference $t-s$, e.g.,
$$
G^{<\bbeta,\bmu,\bv}_+(t-s;x,y)=\lim_{t_0\to-\infty}G^{<}(t,s;x,y)=
\i\langle a^\ast_y\tau_{K_\bv}^{t-s}(a_x)\rangle_+^{\bbeta,\bmu,\bv}.
$$
Accordingly, the steady state density and currents are given by
$$
{\frak n}^+(x)=\Im G^{<\bbeta,\bmu,\bv}_+(0;x,x),\qquad
{\frak j}_j^+=-2d_j\Re\sum_{x\in\cS}G^{<\bbeta,\bmu,\bv}_+(0;0_j,x)\phi_j(x).
$$
We shall denote the Fourier transforms of the NESS Green's functions by 
$\widehat{G}^{\square\bbeta,\bmu,\bv}_+(\omega;x,y)$ so that
$$
{G}^{\square\bbeta,\bmu,\bv}_+(t;x,y)
=\frac1{2\pi}\int_{-\infty}^\infty 
\widehat{G}^{\square\bbeta,\bmu,\bv}_+(\omega;x,y)\e^{\i t\omega}\d\omega.
$$
We note that, a priori, $\widehat{G}^{\square\bbeta,\bmu,\bv}_+(\,\cdot\,;x,y)$ is only defined as a
distribution. In particular $\widehat{G}^{{\rm r/a}\,\bbeta,\bmu,\bv}_+(\omega;x,y)$ is the boundary 
value of an analytic functions on the upper/lower half-plane. Let us briefly explain how these
distributions relate to spectral measures of a self-adjoint operator.

Let $(\cH_+,\Omega_+,\pi_+)$ denote the GNS representation of $\CAR(\fh)$
induced by the $(\bbeta,\bmu,\bv)$-NESS.
Let $L_+$ be the standard Liouvillian of the dynamics $\tau_{K_\bv}$, i.e., the unique 
self-adjoint operator on $\cH_+$ such that 
$\e^{\i tL_+}\pi_+(A)\e^{-\i tL_+}=\pi_+(\tau_{K_\bv}^t(A))$ for all $A\in\CAR(\fh)$
and $L_+\Omega_+=0$ (see, e.g., \cite{AJPP1,Pi}). It immediately follows that
\begin{align*}
G^{<\bbeta,\bmu,\bv}_+(t;x,y)=\i\langle a^\ast_y\tau_{K_\bv}^t(a_x)\rangle_+
&=\i(\Omega_+|\pi_+(a^\ast_y)\e^{\i tL_+}\pi_+(a_x)\e^{-\i tL_+}\Omega_+)\\
&=\i(\pi_+(a_y)\Omega_+|\e^{\i tL_+}\pi_+(a_x)\Omega_+),
\end{align*}
and we conclude that the lesser Green's function $G^{<\bbeta,\bmu,\bv}_+(\,\cdot\,;x,y)$ is essentially 
the Fourier transform of the spectral measure of $L_+$ for the vectors $\pi_+(a_x)\Omega_+$
and $\pi_+(a_y)\Omega_+$. A similar result holds for the greater Green's function.
A simple calculation shows that the Fourier transform of the retarded/advanced Green's functions
can be expressed in terms of the boundary value of the
Borel transform of spectral measures of $L_+$. We note however that since the GNS representation 
of interacting Fermi systems is usually not explicitly known, these relations can hardly be 
exploited (see Section \ref{sct-spectral} for more concrete realizations).

\subsubsection{NESS and scattering theory}
\label{Sct-Moller}

In the absence of electron--electron interactions ($\xi=0$) the well known Landauer-B\"uttiker 
formalism applies and the steady state currents ${\frak j}^+_j$ can be expressed in terms of 
scattering data (see Remark 1 in Section \ref{sct-part} and, e.g., \cite{Im} for a physical introduction). 
In fact, it is possible to relate the NESS $\st_+^{\bbeta,\bmu,\bv}$ 
to the M\o ller operator intertwining the one-particle dynamics of the decoupled system 
$\e^{-\i th_{{\rm D},\bv}}$ to that of the coupled one $\e^{-\i th_\bv}$ (see Equ.\;\eqref{nonNESS}
below). Recently,  several rigorous 
proofs of the Landauer-B\"uttiker formula have been obtained on the basis on this scattering 
approach to the construction of the NESS \cite{AJPP2,CJM,Ne}.

As advocated by Ruelle in \cite{Ru1,Ru2}, the scattering theory of groups of $C^\ast$-algebra
automorphisms (the algebraic counterpart of the familiar Hilbert space scattering theory)
provides a powerful tool for the analysis of weakly interacting many body systems. 
As far as we know, the use of algebraic scattering in this context
can be traced back to the analysis of the s--d model of the Kondo effect by Hepp \cite{He1} 
(see also \cite{He2}). It was subsequently used by Robinson \cite{Ro} to discuss return to equilibrium
in quantum statistical mechanics. More recently, it was effectively applied to the construction
of the NESS of partitioned interacting Fermi gases and to the study of their properties \cite{DFG,FMU,FMSU,JOP3}. 
Let us briefly explain the main ideas behind this approach (we refer the reader to \cite{AJPP1} 
for a detailed pedagogical exposition).

Assuming that at the initial time $t_0$ the system is in a state $\st$ which 
is invariant under the decoupled and non-interacting dynamics $\tau_{H_{{\rm D},\bv}}$
we can write the expectation value of an observable $A\in\CAR(\fh)$ at time $t$ as
$$
\langle\tau_{K_\bv}^{t-t_0}(A)\rangle
=\langle\tau_{H_{{\rm D},\bv}}^{t_0}\circ\tau_{K_\bv}^{-t_0}(\tau_{K_\bv}^{t}(A))\rangle.
$$
If we further assume that for all $A\in\CAR(\fh)$ the limit
\begin{equation}
\varsigma(A)=\lim_{t_0\to-\infty}\tau_{H_{{\rm D},\bv}}^{t_0}\circ\tau_{K_\bv}^{-t_0}(A),
\label{gammadef}
\end{equation}
exists in the norm of $\CAR(\fh)$ then we obtain the following expression for the NESS
$$
\langle\tau_{K_\bv}^t(A)\rangle_+=\lim_{t_0\to-\infty}\langle\tau_{K_\bv}^{t-t_0}(A)\rangle
=\langle\varsigma(\tau_{K_\bv}^{t}(A))\rangle.
$$
The map $\varsigma$ defined by Equ.\;\eqref{gammadef} is an isometric $\ast$-endomorphism of 
$\CAR(\fh)$ which intertwines the two groups $\tau_{H_{{\rm D},\bv}}$ and $\tau_{K_\bv}$, i.e.,
$\tau_{H_{{\rm D},\bv}}^t\circ\varsigma=\varsigma\circ\tau_{K_\bv}^t$. Since it plays a similar 
r\^ole than the familiar M\o ller (or wave) operator of Hilbert space scattering theory,
it is called M\o ller morphism.

To construct the M\o ller morphism $\varsigma$ and hence the NESS $\st_+$ 
we shall invoke the usual chain rule of scattering theory and write $\varsigma$ as the composition 
of two M\o ller morphisms,
$$
\varsigma=\gamma_{\omega_\bv}\circ\varsigma_{\bv},
$$ 
where $\gamma_{\omega_\bv}$ intertwines the decoupled non-interacting dynamics 
$\tau_{H_{{\rm D},\bv}}$ and the coupled non-interacting dynamics $\tau_{H_\bv}$
and $\varsigma_\bv$ intertwines $\tau_{H_\bv}$ with the coupled and interacting dynamics
$\tau_{K_\bv}$. Since  $\gamma_{\omega_\bv}$ does not involve the interaction $W$
it can be constructed by a simple one-particle Hilbert space analysis. The construction of
$\varsigma_\bv$ is more delicate and requires a control of the Dyson expansion
\beq
\tau_{H_\bv}^{-t}\circ\tau_{K_\bv}^t(A)=A+
\sum_{n=1}^\infty\xi^n\!\!\!\!\int\limits_{0\le s_1\le\cdots\le s_n\le t}\!\!\!\!
\i[\tau_{H_\bv}^{-s_n}(W),\i[\tau_{H_\bv}^{-s_{n-1}}(W),\i[\cdots,
\i[\tau_{H_\bv}^{-s_1}(W),A]\cdots]]]
\d s_1\cdots\d s_n.
\label{DysonStart}
\eeq
uniformly in $t$ up to $t=+\infty$. Such a control is possible thanks to the dispersive properties of
the non-interacting dynamics $\tau_{H_\bv}$. We shall rely on the results obtained in \cite{JOP3} 
on the basis of the previous works \cite{BM,Ev,BMa} (a similar analysis can be found in \cite{FMU}).

{\bf Remark.} A serious drawback of this strategy is the fact that the above mentioned
uniform control of the Dyson expansion fails as soon as a bound state occurs in the
coupled non-interacting dynamics, i.e., when the one-body Hamiltonian $h_\bv$ acquires an
eigenvalue. Moreover, the presently available techniques do not allow us to exploit the
repulsive nature of the electron--electron interaction. These are two main reasons which
restrict the analysis to weakly interacting systems (small values of $|\xi|$).

\section{Results}\label{sectiunea3}

To formulate our main assumption, let us define
$$
v_\cS(E,\bv)=
-\lim_{\epsilon\downarrow0}1_\cS\,h_{\rm T}(h_\cR+v_\cR-E-\i\epsilon)^{-1}h_{\rm T}1_\cS,
$$
for $E\in\rr$, $\bv=(v_1,\ldots,v_m)\in\rr^m$ and $v_\cR=\oplus_{j=1}^mv_j 1_j$
(see Equ.\;\eqref{vS} below for a more explicit formula). The following condition ensures that the 
one-particle Hamiltonian $h_\bv$ has neither an eigenvalue nor a real resonance.
\newcommand{\SPv}{{\hyperref[SPv]{{\rm(${\bf SP}_\bv$)}}}}
\newcommand{\SPzero}{{\hyperref[SPv]{{\rm(${\bf SP}_\b0$)}}}}
\begin{quote}\label{SPv}
(${\bf SP}_\bv$) The matrix
\begin{equation}
{\mathfrak m}_\bv(E)=(h_\cS+v_\cS(E,\bv)-E)^{-1},
\label{edef}
\end{equation}
exists for all $E\in\rr$.
\end{quote}
In particular, this implies that $h_\bv$ has purely absolutely continuous spectrum (see Lemma 
\ref{LocDec} below). We note that condition (${\rm SP}_\bv$) imposes severe restrictions on the model.
Indeed, since $h_{D,\bv}$ is bounded, the Hamiltonian $h_\bv$ necessarily acquires eigenvalues 
as the tunneling strength $\max_j|d_j|$ increases. Moreover, if the system $\cS$
is not completely resonant with the leads, i.e., if $h_{D,\bv}$ has non-empty discrete spectrum,
then $h_\bv$ will have eigenvalues at small tunneling strength.

By the Kato-Rosenblum theorem (see e.g., \cite{RS3}), Condition (${\rm SP}_\bv$) implies that
the M\o ller operator
\begin{equation}
\omega_\bv=\slim_{t\to\infty}\e^{-\i th_{{\rm D},\bv}}\e^{\i th_\bv},
\label{omegaplus}
\end{equation}
exists and is complete. Since the absolutely continuous subspace of the decoupled Hamiltonian
$h_{{\rm D},\bv}$ is $\fh_\cR$, 
$\omega_\bv$ is unitary as a map from $\fh$ to $\fh_\cR$. The associated Bogoliubov map 
$\gamma_{\omega_\bv}$, characterized by $\gamma_{\omega_\bv}(a^\#(f))=a^\#(\omega_\bv f)$,
is a $\ast$-isomorphism from $\CAR(\fh)$ to $\CAR(\fh_\cR)$ . Since this map is going to play an
important role in the following, we introduce the short notation
$$
A_\bv=\gamma_{\omega_\bv}(A).
$$
We denote $\tau_{\cR,\bv}$ the quasi-free dynamics on 
$\CAR(\fh)$ generated by the Hamiltonian $h_{\cR,\bv}$ and note that this dynamics has a natural 
restriction to $\CAR(\fh_\cR)$ for which we use the same notation. Let ${\mathcal D_\bv}$ be the linear 
span of $\{\e^{\i th_\bv}f\,|\,t\in\rr, f\in\fh\text{ finitely supported}\}$, a dense subspace of $\fh$.
We denote by ${\mathcal A}_\bv$ the set of polynomials in $\{a^\#(f)\,|\,f\in{\mathcal D}_\bv\}$.
Finally, we set
$$
\Delta_n=\{(s_1,\ldots,s_n)\in\rr^n\,|\,0\le s_1\le s_2\le\cdots \le s_n\}.
$$

\subsection{Existence of the NESS}
\label{sct-part}

Our first result concerns the existence of the $(\bbeta,\bmu,\bv)$-NESS.  It is based on,
and provides extensions of prior results in \cite{FMU,JOP3,AJPP2,Ne}.

\begin{theorem}\label{pthm}
Under Condition \SPv{} there exists a constant $\bar\xi_\bv>0$ such that
the following statements hold if $|\xi|<\bar\xi_\bv$.
\begin{enumerate}[{\rm (1)}]
\item  The limit \eqref{gammadef} exists in the norm of $\CAR(\fh)$ for 
any $A\in\CAR(\fh)$ and defines a $\ast$-isomorphism $\varsigma$ from $\CAR(\fh)$ 
onto $\CAR(\fh_\cR)$.
\item For any $\bbeta\in\rr_+^m$, $\bmu\in\rr^m$ the $(\bbeta,\bmu,\bv)$-NESS exists
and is given by 
$$
\CAR(\fh)\ni A\mapsto
\langle A\rangle^{\bbeta,\bmu,\bv}_+=\langle\varsigma(A)\rangle_\cR^{\bbeta,\bmu}.
$$
\item For $A\in{\mathcal A}_\bv$ the Dyson expansion \eqref{DysonStart} converges up to $t=+\infty$ 
and provides the following convergent power series expansion of the NESS expectation
\beq
\langle A\rangle^{\bbeta,\bmu,\bv}_+
=\langle A_\bv\rangle_\cR^{\bbeta,\bmu}
+\sum_{n=1}^\infty\xi^n\int\limits_{\Delta_n}
C_n(A_\bv;s_1,\ldots,s_n)
\d s_1\cdots\d s_n,
\label{pNESS}
\eeq
where
$$
C_n(A_\bv;s_1,\ldots,s_n)=
\left<
\i[\tau_{\cR,\bv}^{-s_n}(W_\bv),\i[\tau_{\cR,\bv}^{-s_{n-1}}(W_\bv),\i[\cdots,
\i[\tau_{\cR,\bv}^{-s_1}(W_\bv),A_\bv]\cdots]]]
\right>_\cR^{\bbeta,\bmu}\in L^1(\Delta_n).
$$
\end{enumerate}
\end{theorem}

\noindent{\bf Remark 1.} In absence of interaction (i.e., for $\xi=0$) Equ.\;\eqref{pNESS} reduces to
\beq
\langle A\rangle_{+,\xi=0}^{\bbeta,\bmu,\bv}
=\langle\gamma_{\omega_\bv}(A)\rangle_\cR^{\bbeta,\bmu},
\label{nonNESS}
\eeq
and extends to all $A\in\CAR(\fh)$ by continuity.
One immediately deduces from Equ.\;\eqref{GIQF} and \eqref{rhoRbetamu} that this is the 
gauge-invariant quasi-free state on $\CAR(\fh)$ with density
$$
\varrho_+^{\bbeta,\bmu,\bv}=\omega_\bv^\ast\varrho_\cR^{\bbeta,\bmu}\omega_\bv.
$$
In this case one can drop Assumption \SPv{} and show that
$$
\lim_{t\to\infty}\frac1t\int_{-t}^0\langle\tau_{H_\bv}^{-s}(A)\rangle\,\d s=
\langle\gamma_{\omega_\bv}(A)\rangle_\cR^{\bbeta,\bmu},
$$
holds for any almost-$(\bbeta,\bmu)$-KMS state $\st$ and all
$A\in\CAR(\fh)$ provided the one-particle Hamiltonian $h_\bv$ has empty singular continuous
spectrum (note however that the time averaging is necessary as soon as $h_\bv$ has non-empty point 
spectrum). This follows, e.g., from Theorem 3.2 in \cite{AJPP2}. Applying Corollary 4.2 of \cite{AJPP2}
we get the following Landauer-B\"uttiker formulae for the mean currents in the NESS
\begin{align}
\langle J_j\rangle_+^{\bbeta,\bmu,\bv}&=\sum_{k=1}^m\int_{I_{j,k}}
T_{jk}(E,\bv)\left[f(\beta_j(E-v_j-\mu_j))-f(\beta_k(E-v_k-\mu_k))\right]\d E\label{JBL}\\
\langle E_j\rangle_+^{\bbeta,\bmu,\bv}&=\sum_{k=1}^m\int_{I_{j,k}}
T_{jk}(E,\bv)\left[f(\beta_j(E-v_j-\mu_j))-f(\beta_k(E-v_k-\mu_k))\right]
(E-v_j)\d E\label{EBL}
\end{align}
where $I_{j,k}=\sp(h_j+v_j)\cap\sp(h_k+v_k)$, $f(x)=(1+\e^x)^{-1}$ and 
$$
T_{jk}(E,\bv)=d_j^2d_k^2r(E-v_j)r(E-v_k)
|\langle\phi_j|{\mathfrak m}_\bv(E)\phi_k\rangle|^2,\qquad 
r(E)=\left[\frac2{\pi c_\cR^2}\left(1-\left(\frac{E}{2c_\cR}\right)^2\right)\right]^{1/2}.
$$
is the transmission probability trough the sample from lead $\cR_k$ to lead $\cR_j$
at energy $E$ for the one-particle Hamiltonian $h_\bv$.

\noindent{\bf Remark 2.} 
The intertwining property of the morphism $\gamma_{\omega_\bv}$ allows us to rewrite
Equ.\;\eqref{pNESS} in term of the non-interacting NESS \eqref{nonNESS} as
$$
\langle A\rangle^{\bbeta,\bmu,\bv}_+=\langle A\rangle_{+,\xi=0}^{\bbeta,\bmu,\bv}
+\sum_{n=1}^\infty\xi^n\!\!\!\int\limits_{\Delta_n}\!\!\!
\left<
\i[\tau_{H_\bv}^{-s_n}(W),\i[\tau_{H_\bv}^{-s_{n-1}}(W),\i[\cdots,\i[\tau_{H_\bv}^{-s_1}(W),A]\cdots]]]
\right>_{+,\xi=0}^{\bbeta,\bmu,\bv}\d s_1\cdots\d s_n,
$$
for $A\in{\mathcal A}_\bv$.

\noindent{\bf Remark 3.} It follows from Theorem \ref{pthm} that for $A\in{\mathcal A}_\bv$
the NESS expectation $\langle A\rangle^{\bbeta,\bmu,\bv}_+$ is an analytic function of the 
interaction strength $\xi$ for $|\xi|<\bar\xi_\bv$. For computational purposes, its Taylor 
expansion around $\xi=0$ can be obtained by iterating the integral equation
$$
\eta_t(A)=A+\xi\int_0^t\i[\tau_{H_\bv}^{-s}(W),\eta_s(A)]\d s,
$$
setting $t=\infty$ in the resulting expression and writing
$$
\langle A\rangle^{\bbeta,\bmu,\bv}_+=\langle \eta_{t=+\infty}(A)\rangle_{+,\xi=0}^{\bbeta,\bmu,\bv}.
$$

\noindent{\bf Remark 4.} The spectrum of the one-body Hamiltonian $h_{\cR,\bv}$ acting on 
$\fh_\cR$ being purely absolutely continuous, the quasi-free $C^\ast$-dynamical system
$(\CAR(\fh_\cR),\tau_{\cR,\bv}^t,\st_{\cR}^{\bbeta,\bmu})$ is mixing
(see, e.g., \cite{JP2,Pi,AJPP1}). Since the $\ast$-isomorphism $\varsigma$ intertwines this system
with the $C^\ast$-dynamical system 
$(\CAR(\fh),\tau_{K_\bv}^t,\st_+^{\bbeta,\bmu,\bv})$,  it follows that the latter 
is also mixing, i.e.,
$$
\lim_{t\to\infty}\langle A\tau_{K_\bv}^t(B)\rangle_+^{\bbeta,\bmu,\bv}
=\langle A\rangle_+^{\bbeta,\bmu,\bv}\langle B\rangle_+^{\bbeta,\bmu,\bv},
$$
holds for all $A,B\in\CAR(\fh)$. It also follows that if the leads are initially near a common equilibrium 
state, i.e., if $\bbeta=(\beta,\ldots,\beta)$ and 
$\bmu=(\mu,\ldots,\mu)$, then the restriction of $\st_{\cR}^{\bbeta,\bmu}$
to $\CAR(\fh_\cR)$ is the unique $(\beta,\mu)$-KMS state for the dynamics $\tau_{\cR}$ and hence
$\st_+^{\bbeta,\bmu,\b0}=\st_{K_\b0}^{\beta,\mu}$ is the unique 
$(\beta,\mu)$-KMS state on $\CAR(\fh)$ for the zero-bias dynamics $\tau_{K_\b0}$.

\noindent{\bf Remark 5.} The linear response theory of the partitioned NESS
$\st^{\bbeta,\bmu,\b0}_+$ was established in \cite{JOP3,JPP}. In particular the
Green-Kubo formula
\begin{equation}
L_{jk}=\frac{1}{\beta}
\left.\partial_{\mu_k}\langle J_j\rangle^{\bbeta,\bmu,\b0}_+
\right|_{\bbeta=(\beta,\ldots,\beta),\bmu=(\mu,\ldots,\mu)}=\frac{1}{\beta}
\lim_{t\to\infty}
\int_{0}^t\left[\int_0^\beta
\langle\tau_{K_\b0}^s(J_j)\tau_{K_\b0}^{\i\theta}(J_k)\rangle_{K_\b0}^{\beta,\mu}\d\theta
\right]\d s,
\label{GKform}
\end{equation}
holds for the charge current observable $J_j$ of Equ.\;\eqref{Jdef}. If the system is time
reversal invariant (see Remark~2, Section~\ref{FermiState}) then this can be rewritten as
\begin{equation*}
L_{jk}=\frac12\int_{-\infty}^\infty
\langle\tau_{K_\b0}^s(J_j)J_k\rangle_{K_\b0}^{\beta,\mu}\d s.
\label{GKform2}
\end{equation*}
The last formula further yields the Onsager reciprocity relation $L_{jk}=L_{kj}$ (see \cite{JOP3,JPP} for 
details. Similar results hold for the energy currents).

\noindent{\bf Remark 6.} To the best of our knowledge, the linear response theory of the
partition-free NESS has not yet been studied. In particular we do not know if the Green-Kubo formula
$$
\left.\partial_{v_k}\langle J_j\rangle^{\bbeta,\bmu,\bv}_+
\right|_{\bbeta=(\beta,\ldots,\beta),\bmu=(\mu,\ldots,\mu),\bv=\b0}
=\lim_{t\to\infty}\int_0^t\left[\int_0^\beta
\langle\tau_{K_\b0}^s(J_j)\tau_{K_\b0}^{\i\theta}(J_k)\rangle_{K_\b0}^{\beta,\mu}\d\theta
\right]\d s,
$$
holds. We note however that this formula can be explicitly 
checked in the non-interacting case with the help of the Landauer-B\"uttiker formula \eqref{JBL}.
Moreover, it easily follows from Duhamel's formula and Lemma 4.7 in \cite{JOP2} that the 
{\sl finite time} Green-Kubo formula
$$
\left.\partial_{v_k}\langle \tau_{K_\bv}^t(J_j)\rangle^{\beta,\mu}_{K_\b0}\right|_{\bv=\b0}
=\int_{0}^t\left[\int_0^\beta
\langle\tau_{K_\b0}^s(J_j)\tau_{K_\b0}^{\i\theta}(J_k)\rangle_{K_\b0}^{\beta,\mu}\d\theta
\right]\d s,
$$
holds for all $t\ge0$. If the system is time reversal invariant then
Proposition 4.1 in \cite{JOP2} and Remark 4 imply that
$$
\lim_{t\to\infty}\left.\partial_{v_k}\langle \tau_{K_\bv}^t(J_j)\rangle^{\beta,\mu}_{K_\b0}
\right|_{\bv=\b0}=
\frac{\beta}{2}\int_{-\infty}^\infty\langle\tau_{K_\b0}^s(J_j)J_k\rangle_{K_\b0}^{\beta,\mu}\d s.
$$
Thus, the Green-Kubo formula holds iff the limit $t\to\infty$ commutes with the partial derivative
$\partial_{v_k}$ at $\bv=0$,
$$
\lim_{t\to\infty}\left.\partial_{v_k}\langle \tau_{K_\bv}^t(J_j)\rangle^{\beta,\mu}_{K_\b0}
\right|_{\bv=\b0}=\left.\partial_{v_k}\lim_{t\to\infty}\langle \tau_{K_\bv}^t(J_j)\rangle^{\beta,\mu}_{K_\b0}
\right|_{\bv=\b0}.
$$

\subsection{Adiabatic schemes}
\label{sct-free}

Our second result shows that an adiabatic switching of the coupling between the sample and the leads
or of the bias does not affect the NESS.

\begin{theorem}\label{pfthm} 
\begin{enumerate}[{\rm (1)}] 
\item Assume that Condition \SPzero{} holds and that $|\xi|<\bar\xi_\b0$.
Then the adiabatic evolution $\alpha_{\b0,t_0}^{s,t}$ satisfies
$$
\lim_{t_0\to-\infty}\langle\alpha_{\b0,t_0}^{t_0,0}(A)\rangle=\langle A\rangle^{\bbeta,\bmu,\b0}_+,
$$
for any $A\in\CAR(\fh)$ and any almost-$(\bbeta,\bmu)$-KMS state $\st$, 
i.e., adiabatic switching of the coupling produces the same NESS as instantaneous coupling.

\item Assume that Condition \SPv{} holds and that $|\xi|<\bar\xi_\bv$. Then
the adiabatic evolution $\alpha_{\bv,{t_0}}^{s,t}$ satisfies
$$
\lim_{t_0\to-\infty}\langle\alpha_{\bv,{t_0}}^{t_0,0}(A)\rangle
=\langle A\rangle^{\bbeta,\bmu,\bv}_+,
$$
for any $A\in\CAR(\fh)$ and any almost-$(\bbeta,\bmu)$-KMS state $\st$, 
i.e., adiabatic switching of the bias produces the same NESS as instantaneous switching.
\end{enumerate}
\end{theorem}

\noindent{\bf Remark 1.} Part (1) is an instance of the Narnhofer-Thirring adiabatic theorem \cite{NT}. 
Our proof is patterned on the proof given in \cite{NT}. Part (2) employs some of the ideas developed
in \cite{CDP}; here the problem is 'easier' due to the absence of point spectrum. 

\subsection{Entropy production}

We denote Araki's relative entropy of two states by ${\rm S}(\,\cdot\,|\,\cdot\,)$ with the notational
convention of \cite{BR2}. The following result establishes the relation between the rate of divergence
of this relative entropy along the flow of the dynamics $\tau_{K_\bv}$ and the phenomenological
notion of entropy production rate of the $(\bbeta,\bmu,\bv)$-NESS.

\begin{theorem}\label{entropicthm}
\begin{enumerate}[{\rm (1)}] 
\item There exists a norm dense subset $\mathfrak S$ of the set of all
almost-$(\bbeta,\bmu)$-KMS states such that, for all $\st\in{\frak S}$ one has
\beq
\lim_{t_0\to-\infty}\frac1{t_0}{\rm S}(\langle\tau_{K_\bv}^{-t_0}(\,\cdot\,)\rangle|\st)
=-\sum_{j=1}^m\langle\beta_j( E_j-\mu_j J_j)\rangle^{\bbeta,\bmu,\bv}_+\ge0.
\label{SecLaw}
\eeq
\item The mean entropy production rate in the partition-free NESS is given by
$$
\lim_{t_0\to-\infty}\frac1{t_0} 
{\rm S}(\langle\tau^{-t_0}_{K_\bv}(\,\cdot\,)\rangle_{K_\b0}^{\beta,\mu}|
\st_{K_\b0}^{\beta,\mu})
=\sum_{j=1}^m\beta v_j\langle J_j\rangle_+^{\beta,\mu,\bv}\ge0.
$$
\item If $\bbeta=(\beta,\ldots,\beta)$, $\bmu=(\mu,\ldots,\mu)$ and $\bv=(v,\ldots,v)$,
then the  $(\bbeta,\bmu,\bv)$-NESS is the unique $(\beta,\mu)$-KMS state for the dynamics 
$\tau_{K_\b0-vN_\cS}$ and all the steady currents vanish 
$$
\langle J_j\rangle_+^{\bbeta,\bmu,\bv}=0,\qquad\langle E_j\rangle_+^{\bbeta,\bmu,\bv}=0.
$$
\end{enumerate}
\end{theorem}

The inequality on the right hand side of Equ.\;\eqref{SecLaw} is
related to the second law of thermodynamics. Indeed, according to phenomenological
thermodynamics, the quantity
$$
-\sum_{j=1}^m\beta_j\left(\langle E_j\rangle_+^{\bbeta,\bmu,\bv}
-\mu_j\langle J_j\rangle^{\bbeta,\bmu,\bv}_+\right),
$$
can be identified with the mean rate of entropy production in the steady state
$\st^{\bbeta,\bmu,\bv}_+$
(see \cite{Ru2,JP2,JP3} for more details). For non-interacting systems, the Landauer-B\"uttiker
formulae \eqref{JBL},\eqref{EBL} yield the following expression of this entropy production rate
$$
-\sum_{j=1}^m\langle\beta_j( E_j-\mu_j J_j)\rangle^{\bbeta,\bmu,\bv}_+=
\sum_{j,k=1}^m\int_{I_{j,k}}
T_{jk}(E,\bv)\left[f(x_k(E))-f(x_j(E))\right]x_j(E)\frac{\d E}{2\pi},
$$
where $x_j(E)=\beta_j(E-v_j-\mu_j)$. As shown in \cite{AJPP2,Ne}, the right hand side of this
identity is strictly positive if there exists a pair $(j,k)$ such that the transmission probability 
$T_{jk}(E,\bv)$ does not vanish identically and either $\beta_j\not=\beta_k$ or 
$v_j+\mu_j\not=v_k+\mu_k$.
The analytic dependence of the NESS expectation on the interaction strength
displayed by Equ.\;\eqref{pNESS} allows us to conclude that this situation persists for sufficiently 
weak interactions (a fact already proved in \cite{FMU}). Strict positivity of entropy
production for weakly interacting fermions out of equilibrium is a generic property, as shown in
\cite{JP5}. In the more general context of open quantum systems it can also be proved in the limit of 
weak coupling to the reservoirs (more precisely in the van Hove scaling limit) which
provides another perturbative  approach to this important problem (see \cite{LS,AS,DM,DRM}).

\subsection{The NESS Green-Keldysh functions}
\label{sct-spectral}

In the next result we collect some important properties of the Green-Keldysh correlation
functions of the $(\bbeta,\bmu,\bv)$-NESS. In particular we relate these functions to the spectral 
measures of an {\sl explicit} self-adjoint operator.

Since the restriction of the state $\st_\cR^{\bbeta,\bmu}$ to $\CAR(\fh_\cR)$ is quasi-free with
density $\varrho_0=\varrho_\cR^{\bbeta,\bmu}|_{\fh_\cR}$, it induces a GNS representation
$(\cH_\cR,\Omega_\cR,\pi_\cR)$ of $\CAR(\fh_\cR)$ of Araki-Wyss type (see \cite{AW,DeGe,AJPP1}).
More precisely, one has
$$
\cH_\cR=\Gamma_-(\fh_\cR)\otimes\Gamma_-(\fh_\cR), \qquad
\Omega_\cR=\Omega\otimes\Omega,
$$
$$
\pi_\cR(a(f))=a\left((I-\rho_0)^{1/2}f\right)\otimes I
+\e^{\i\pi N}\otimes a^\ast\left(\overline{\rho_0^{1/2}f}\right),
$$
where $\Omega\in\Gamma_-(\fh_\cR)$ denotes the Fock vacuum vector, $N=\d\Gamma(I)$ is
the number operator and $\overline{\vphantom{f}\,\cdot\,}$ denotes the usual complex conjugation on 
$\fh_\cR=\oplus_{j=1}^m\ell^2(\nn)$. The standard Liouvillian 
$$
L_{\cR,\bv}=\d\Gamma(h_{\cR,\bv})\otimes I-I\otimes\d\Gamma(h_{\cR,\bv}),
$$
is the unique self-adjoint operator on $\cH_\cR$ such that 
$\e^{\i tL_{\cR,\bv}}\pi_\cR(A)\e^{-\i tL_{\cR,\bv}}=\pi_\cR(\tau_{\cR,\bv}^t(A))$
for all $A\in\CAR(\fh_\cR)$ and $L_{\cR,\bv}\Omega_\cR=0$. Apart for this eigenvector, the
Liouvillian $L_{\cR,\bv}$ has a purely absolutely continuous spectrum filling the entire real axis.

\begin{theorem}\label{specthm}
Assume that Condition \SPv{} holds. Then the series
$$
A_x=
a(\omega_\bv\delta_x)+\sum_{n=1}^\infty\xi^n\int_{\Delta_n}
\i[\tau_{\cR,\bv}^{-s_n}(W_\bv),\i[\tau_{\cR,\bv}^{-s_{n-1}}(W_\bv),\i[\cdots,\i[\tau_{\cR,\bv}^{-s_1}(W_\bv),
a(\omega_\bv\delta_x)]\cdots]]]\d s_1\cdots\d s_n,
$$
is norm convergent for $|\xi|<\bar\xi_v$ and defines an element of $\CAR(\fh_\cR)$. For
$x\in\cS\cup\cR$, set
$$
\Psi_x=\pi_\cR(A_x)\Omega_\cR,\qquad\Psi_x^\ast=\pi_\cR(A_x)^\ast\Omega_\cR,
$$
and denote by $\lambda_{\Phi,\Psi}$ the spectral measure of $L_{\cR,\bv}$ for $\Phi$ and $\Psi$. 
Then, for any $x,y\in\cS\cup\cR$, one has:
\begin{enumerate}[{\rm (1)}] 
\item $\Psi_x$ and $\Psi^\ast_x$ are orthogonal to $\Omega_\cR$.
\item The complex measure $\lambda_{\Psi_y,\Psi_x}$ is absolutely continuous w.r.t.\;Lebesgue's
measure and
$$
\widehat{G}^{<\bbeta,\bmu,\bv}_+(\omega;x,y)
=2\pi\i\frac{\d\lambda_{\Psi_y,\Psi_x}(\omega)}{\d\omega}.
$$
\item The complex measure $\lambda_{\Psi_x^\ast,\Psi_y^\ast}$ is absolutely continuous 
w.r.t.\;Lebesgue's measure and
$$
\widehat{G}^{>\bbeta,\bmu,\bv}_+(\omega;x,y)
=-2\pi\i\frac{\d\lambda_{\Psi_x^\ast,\Psi_y^\ast}(-\omega)}{\d\omega}.
$$
\end{enumerate}
In the remaining statements, $\square$ stands for either $<$ or $>$.
\begin{enumerate}[{\rm (1)}] 
\setcounter{enumi}{3}
\item $G^{\square\bbeta,\bmu,\bv}_+(t;x,y)$
extends to an entire analytic functions of $t$. Moreover, for all $n\ge0$ and $z\in\cc$,
$$
\lim_{t\to\pm\infty}\partial_z^nG^{\square\bbeta,\bmu,\bv}_+(z+t;x,y)=0.
$$
\item There exists $\theta>0$ such that
$$
\sup_{|\Im z|\le\theta}\int_{-\infty}^\infty\left|G^{\square\bbeta,\bmu,\bv}_+(z+t;x,y)\right|\d t<\infty.
$$
for all $x,y\in\cS\cup\cR$.
\item $\widehat{G}^{\square\bbeta,\bmu,\bv}_+(\omega;x,y)$ is a continuous function of $\omega$. 
Moreover, 
$$
\sup_{\omega\in\rr}\e^{\theta|\omega|}
\left|\widehat{G}^{\square\bbeta,\bmu,\bv}_+(\omega;x,y)\right|<\infty,
$$
holds for all $x,y\in\cS\cup\cR$ with the same $\theta$ as in Part (5). 
\end{enumerate}
\end{theorem}

\subsection{The Hartree-Fock approximation}
\label{sct-first}

In this section we focus on two-body interactions $W$ as given by Equ.\;\eqref{Vint}. 
Recall that $w(x,x)=0$ for all $x\in\cS$ and define the Hartree and  exchange interactions by
\begin{align*}
W_{\rm H}&=\sum_{x,y\in\cS}w(x,y)
\langle a^\ast_ya_y\rangle_{+,\xi=0}^{\bbeta,\bmu,\bv}\,a^\ast_xa_x
=\d\Gamma(v_{\rm H}),\\
W_{\rm X}&=\sum_{x,y\in\cS}w(x,y)
\langle a^\ast_ya_x\rangle_{+,\xi=0}^{\bbeta,\bmu,\bv}\,a^\ast_xa_y
=\d\Gamma(v_{\rm X}),
\end{align*}
and the Hartree-Fock interaction $W_{\rm HF}=W_{\rm H}-W_{\rm X}=\d\Gamma(v_{\rm HF})$.
Recall also that $\st_{+,\xi=0}^{\bbeta,\bmu,\bv}$ denotes the non-interacting NESS given by
Equ.\;\eqref{nonNESS}. With the notation of Remark 1 in Section \ref{sct-part} one has
for $x,y\in\cS$,
\begin{align*}
\langle a^\ast_ya_x\rangle_{+,\xi=0}^{\bbeta,\bmu,\bv}
&=\langle\delta_x|\omega_\bv^\ast\varrho_\cR^{\bbeta,\bmu}\omega_\bv\delta_y\rangle\\
&=\sum_{j=1}^m d_j^2 \int_{v_j-2c_\cR}^{v_j+2c_\cR} 
f(\beta_j(E-v_j-\mu_j)) r(E-v_j)\langle\delta_x|{\mathfrak m}_\bv(E)\phi_j\rangle
\langle\phi_j|{\mathfrak m}_\bv(E)^\ast\delta_y\rangle\,\frac{\d E}{\sqrt{2\pi}}.
\end{align*}
The one-particle Hartree-Fock Hamiltonian $h_{{\rm HF},\bv}=h_\bv+\xi v_{\rm HF}$ generates 
a quasi-free dynamics $\tau_{H_{{\rm HF},\bv}}$ on $\CAR(\fh)$.

Our last result shows that the Green-Keldysh correlation functions of the Hartree-Fock dynamics 
$G_{{\rm HF}+}^{\square\bbeta,\bmu,\bv}$ coincide with the one of the fully interactinng
$(\bbeta,\bmu,\bv)$-NESS to first order in the interaction strength.

\begin{theorem}\label{HF-thm}
Assume that Condition  \SPv{} holds.
\begin{enumerate}[{\rm (1)}] 
\item If $\xi$ is small enough then the limit
$$
\langle A\rangle_{{\rm HF}+}^{\bbeta,\bmu,\bv}=
\lim_{t_0\to-\infty}\langle\tau_{H_{{\rm HF},\bv}}^{-t_0}(A)\rangle,
$$
exists for all $A\in\CAR(\fh)$. Moreover this Hartree-Fock $(\bbeta,\bmu,\bv)$-NESS is given by
$\langle A\rangle_{{\rm HF}+}^{\bbeta,\bmu,\bv}=\langle\gamma_{\omega_{{\rm HF},\bv}}(A)\rangle_\cR^{\bbeta,\bmu}$
where
$$
\omega_{{\rm HF},\bv}=\slim_{t\to\infty}\e^{-\i t h_{{\rm D},\bv}}\e^{i t h_{{\rm HF},\bv}}.
$$
\item Denote by $G_{{\rm HF}+}^{\square\bbeta,\bmu,\bv}(t;x,y)$ the Green-Keldysh correlation 
functions of the Hartree-Fock NESS. Then
\beq
G_+^{\square}(t;x,y)=G_{{\rm HF}+}^{\square}(t;x,y)+{\mathcal O}(\xi^2)
\label{HFform}
\eeq
as $\xi\to0$. Moreover, the error term is locally uniform in $x$, $y$ and $t$.
\end{enumerate}
\end{theorem}

Since the NESS expectation of the energy and charge currents can be expressed in terms of
the lesser Green-Keldysh function, one has in particular
$$
\langle J_j\rangle_+^{\bbeta,\bmu,\bv}
=\langle J_j\rangle_{{\rm HF}+}^{\bbeta,\bmu,\bv}+{\mathcal O}(\xi^2),\qquad
\langle E_j\rangle_+^{\bbeta,\bmu,\bv}
=\langle E_j\rangle_{{\rm HF}+}^{\bbeta,\bmu,\bv}+{\mathcal O}(\xi^2).
$$
Moreover, the Hartree-Fock steady currents are given by the Landauer-B\"uttiker formulae
\eqref{JBL}, \eqref{EBL} with the transmission probability
$$
T_{jk}(E,\bv)=d_j^2d_k^2r(E-v_j)r(E-v_k)
|\langle\phi_j|{\mathfrak m}_{{\rm HF},\bv(E)}\phi_k\rangle|^2,
$$
where ${\mathfrak m}_{{\rm HF},\bv(E)}=(h_\cS+\xi v_{\rm HF}+v_\cS(E,\bv)-E)^{-1}$.

\section{Proofs}
\label{sectiunea4}

\subsection{Preliminaries}
\label{presect}

In this section we state and prove a few technical facts which will be used in the proof of our main 
results. We start with some notation. Recall that $1_\cS$, $1_\cR$ and $1_j$ denote the orthogonal 
projections on $\fh_\cS$, $\fh_\cR$ and $\fh_j$ in $\fh$. We set $x=\oplus_{j=1}^mx_j$ where $x_j$
is the position operator on lead $\cR_j$ and use the convention 
$\jx=(1+|x|)$. Thus, the operator $\jx$ acts as the identity on $\fh_\cS$ and as
$(1+x_j)$ on $\fh_j$. In particular, one has
$h_{\rm T}=\jx^\sigma h_{\rm T}=h_{\rm T}\jx^\sigma$ for arbitrary $\sigma\in\rr$.

\begin{lemma}\label{LocDec}
If Condition \SPv{} is satisfied then the one particle Hamiltonian $h_\bv$ has purely absolutely 
continuous spectrum. Moreover, for $\sigma>5/2$ they are constants $C_\sigma$ and $c_\sigma$
such that the local decay estimates
\begin{align}
\|\jx^{-\sigma}\e^{\i(t+\i\theta)h_\bv}\jx^{-\sigma}\|
&\le C_\sigma\e^{c_\sigma|\theta|}\langle t\rangle^{-3/2},\label{lde}\\[6pt]
\|\jx^{-\sigma}\varrho_+^{\bbeta,\bmu,\bv}\e^{\i(t+\i\theta)h_\bv}\jx^{-\sigma}\|
&\le C_\sigma\e^{c_\sigma|\theta|}\langle t\rangle^{-3/2}\label{lderho},
\end{align}
hold for all $t,\theta\in\rr$.
\end{lemma}

\noindent{\bf Proof.} Define
$$
d_\cR(z)=1_\cR\,\jx^{-\sigma}(h_\cR+v_\cR-z)^{-1}\jx^{-\sigma}1_\cR,
$$
for $\Im z\not=0$. The explicit formula for the resolvent of the Dirichlet Laplacian on $\nn$
yields, for $x,y\in\cR_j$, $\theta\in[0,\pi]$ and $\chi>0$,
\begin{equation}
\begin{split}
\lim_{\epsilon\downarrow0}
\langle\delta_x|(h_j-2c_\cR\cos\theta-\i\epsilon)^{-1}\delta_y\rangle&=
-\frac{\e^{-\i\theta|x-y|}-\e^{-\i\theta(x+y+2)}}{2\i c_\cR\sin\theta},\\[6pt]
\langle\delta_x|(h_j\mp 2c_\cR\cosh\chi)^{-1}\delta_y\rangle&=
-\frac{\e^{-\chi|x-y|}-\e^{-\chi(x+y+2)}}{2c_\cR\sinh\chi}(\pm1)^{x+y+1}.
\end{split}
\label{ResForm}
\end{equation}
For $\sigma>1/2$, it follows that the function $d_\cR(z)$, taking values in the Hilbert-Schmidt operators
on $\fh_\cR$, has boundary values $d_\cR(E)=\lim_{\epsilon\downarrow0}d_\cR(E+\i\epsilon)$ at 
every point $E\in\rr$ and $v_\cS(E,\bv)=h_{\rm T}d_\cR(E)h_{\rm T}$.
Moreover, a simple calculation shows that if $\sigma>5/2$, then the following
holds:
\begin{enumerate}[{\rm (1)}]
\item $d_\cR(E)$ is continuous on $\rr$.
\item $d_\cR(E)$ is twice continuously differentiable on $\rr\setminus\mathcal{T}$
where $\mathcal{T}=\{v_j\pm2c_\cR\,|\,j=1,\ldots,m\}$ is the set of thresholds of 
$h_\cR+v_\cR$.
\item For $k=1,2$,
$$ 
\partial_E^k\,d_\cR(E)=\mathcal{O}(\delta(E)^{-k+1/2}),
$$
as $\delta(E)\to0$, where $\delta(E)={\rm dist}(E,\mathcal{T})$.
\end{enumerate}
Note that the class of operator valued functions satisfying Conditions (1)--(3) form an algebra.

The Feshbach-Schur formula and Condition \SPv{} imply that for every $E\in\rr$ the weighted 
resolvent of $h_\bv$ has a boundary value  
$d(E)=\lim_{\epsilon\downarrow0}\jx^{-\sigma}(h_\bv-E-\i\epsilon)^{-1}\jx^{-\sigma}$ given by
$$
d(E)=
\left[
\begin{array}{cc}
{\mathfrak m}_\bv(E)&-{\mathfrak m}_\bv(E)h_{\rm T} d_\cR(E)\\[4pt]
-d_\cR(E)h_{\rm T}{\mathfrak m}_\bv(E)&d_\cR(E)+d_\cR(E)h_{\rm T}{\mathfrak m}_\bv(E)h_{\rm T}d_\cR(E)
\end{array}
\right].
$$
From these expressions, one concludes that $d(E)$ also satisfies the above properties (1)--(3).
In particular, this implies that $h_\bv$ has purely absolutely continuous spectrum.

Denote by $\Sigma_1,\Sigma_2,\ldots$ the connected components of the bounded open set
$\cup_k]v_k-2c_\cR,v_k+2c_\cR[\setminus\mathcal{T}$. By Cauchy's formula one has
$$
\jx^{-\sigma}\e^{\i th_\bv}\jx^{-\sigma}=\frac1\pi\sum_k\int_{\Sigma_k} \e^{\i tE}\,\Im d(E)\d E,
$$
for $t\in\rr$ and the local decay estimate \eqref{lde} with $\theta=0$ follows from Lemma 10.2 in 
\cite{JK}. 

To prove \eqref{lderho} note that $\varrho_+^{\bbeta,\bmu,\bv}\e^{\i th_\bv}
=\sum_j\omega_\bv^\ast 1_jf_j(h_j)\e^{\i t(h_j+v_j)}1_j\omega_\bv$ where 
$f_j(E)=(1+\e^{\beta_j(E-\mu_j)})^{-1}$. Let $U_j$ denote the unitary map from $\fh_j$ 
to the spectral representation of $h_j+v_j$ in $L^2([v_j-2c_\cR,v_j+2c_\cR],\d E)$. From the
stationary representation of the M\o ller operator (see, e.g., \cite{P}) we deduce
$$
\left(U_j1_j(\omega_\bv-I)\jx^{-\sigma}g\right)(E)
=-\left(U_j1_jh_{\rm T}d(E)^\ast g\right)(E),
$$
which implies
$$
\jx^{-\sigma}\varrho_+^{\bbeta,\bmu,\bv}\e^{\i th_\bv}\jx^{-\sigma}=\frac1\pi\sum_{j=1}^m
\sum_k\int_{\Sigma_k}\e^{\i tE}f_j(E-v_j)\left(I-d(E)h_{\rm T}\right)
1_j\Im(d_\cR(E))1_j\left(I-h_{\rm T}d(E)^\ast\right)\d E.
$$
Applying again Lemma 10.2 of \cite{JK} yields \eqref{lderho} with $\theta=0$.

To extend our estimates to non-zero $\theta$, we write
$$
\e^{\i(t+\i\theta)h_\bv}\jx^{-\sigma}
=\e^{\i th_\bv}\jx^{-\sigma}\left(\jx^{\sigma}\e^{-\theta h_\bv}\jx^{-\sigma}\right),
$$
and note that
$$
\|\jx^{\sigma}\e^{-\theta h_\bv}\jx^{-\sigma}\|=\|\e^{-\theta\jx^{\sigma}h_\bv\jx^{-\sigma}}\|
\le\e^{|\theta|\,\|\jx^{\sigma}h_\bv\jx^{-\sigma}\|}.
$$
The desired result follows from the fact that 
$\jx^{\sigma}h_\bv\jx^{-\sigma}=\jx^{\sigma}h_\cR\jx^{-\sigma}+h_\cS+h_{\rm T}+v_\cR$ and
the simple bound $\|\jx^{\sigma}h_\cR\jx^{-\sigma}\|\le(1+4^\sigma)c_\cR$ which follows 
from an explicit calculation.\hfill$\square$

\bigskip
\noindent{\bf Remark 1.} It follows from this proof that 
$\|\jx^\sigma\e^{\i th_\bv}f\|\le\|\jx^\sigma\e^{\i th_\bv}\jx^{-\sigma}\|\,\|\jx^\sigma f\|
\le\e^{c_\sigma|t|}\|\jx^\sigma f\|$ so that the subspace ${\mathcal D}_\bv$ belongs to
the domain of $\jx^\sigma$ for all $\sigma$.

\noindent{\bf Remark 2.} In order to check Condition \SPv{} it may be useful to note that
\begin{equation}
v_\cS(E,\bv)=
\sum_{j=1}^m d_j^2 g_j(E,\bv)\,|\phi_j\rangle\langle\phi_j|,
\label{vS}
\end{equation}
where, according to \eqref{ResForm},
\begin{equation}
g_j(E,\bv)=-\lim_{\epsilon\downarrow0}\,
\langle\delta_{0_j}|(h_j+v_j-E-\i\epsilon)^{-1}\delta_{0_j}\rangle
=\left\{
\begin{array}{ll}
\ds\frac1{c_\cR}\e^{-\i\theta},&\text{ for } E=v_j+2c_\cR\cos\theta, \theta\in[0,\pi];\\[10pt]
\ds\pm\frac1{c_\cR}\e^{-\chi},&\text{ for } E=v_j\pm2c_\cR\cosh\chi, \chi\ge0.
\end{array}
\right.
\label{gForm}
\end{equation}

\begin{lemma}\label{localautonomy}
For any $s,t\in\rr$, $\bv\in\rr^m$ and $A\in\CAR(\fh)$ one has
$$
\lim_{t_0\to-\infty}\|\alpha_{\bv,t_0}^{s,t}(A)-\tau_{K_\bv}^{t-s}(A)\|=0.
$$
\end{lemma}

\noindent{\bf Proof.} 
We first observe that it suffices to prove the claim for the special case $A=a(f)$ for $f\in\fh$.
Since
$K_\bv(t/t_0)-K_\bv=-(1-\chi(t/t_0))\d\Gamma(b_\bv)$, with $b_\b0=h_{\rm T}$  and
$b_\bv=v_\cR$ for $\bv\not=\b0$, Duhamel's formula yields
\begin{align*}
\alpha_{\bv,t_0}^{s,t}(a(f))-\tau_{K_\bv}^{t-s}(a(f))&=-\int_s^t \alpha_{\bv,t_0}^{s,u}
\left(\i[\d\Gamma(b_\bv),\tau_{K_\bv}^{t-u}(a(f))]\right)(1-\chi(u/t_0))\d u\\[4pt]
&=-\int_s^t \alpha_{\bv,t_0}^{s,u}
\left(\i[\d\Gamma(b_\bv),\Gamma_\bv^{t-u}\tau_{H_\bv}^{t-u}(a(f))\Gamma_\bv^{(t-u)\ast}]
\right)(1-\chi(u/t_0))\d u,
\end{align*}
where $\Gamma_\bv^t=\e^{\i tK_\bv}\e^{-\i tH_\bv}$. One easily shows that
$$
[\d\Gamma(b_\bv),\Gamma_\bv^t]
=\i\xi\int_0^t\e^{\i sK_\bv}[\d\Gamma(\e^{-\i sh_\bv}b_\bv\e^{\i sh_\bv}),W]
\e^{\i(t-s)K_\bv}\e^{-\i tH_\bv}\d s.
$$
Since $W$ is a polynomial in $a^\#(f)$, with $f\in\fh_\cS$, there exists a constant $C_W$
such that $\|[\d\Gamma(\e^{-\i sh_\bv}b_\bv\e^{\i sh_\bv}),W]\|\le C_W\|b_\bv\|$ and hence we
have the bound
$$
\|[\d\Gamma(b_\bv),\Gamma_\bv^t]\|\le C_W\|b_\bv\||t\xi|.
$$
This yields the estimate
$$
\left\|\alpha_{\bv,t_0}^{s,t}(a(f))-\tau_{K_\bv}^{t-s}(a(f))\right\|\le
\left(1+2C_W|(t-s)\xi|\right)\|b_\bv\|\,\|f\|
\int_s^t(1-\chi(u/t_0))\d u,
$$
and since $\chi:\rr\to[0,1]$ is continuous with $\chi(0)=1$ the result follows from the dominated 
convergence theorem.

\hfill$\square$

\begin{lemma}\label{asymptoticabelian}
Assume that Condition \SPv{} is satisfied. If $A\in\CARv(\fh)$ and $B\in\CAR(\fh)$
are polynomials with factors in $\{a^\#(f)\,|\,f\in\mathcal{D}_\bv\}$, then
$$
\int_{-\infty}^\infty\|[A,\tau_{H_\bv}^t(B)]\|\d t<\infty.
$$
\end{lemma}

\noindent{\bf Proof.}
The result is a direct consequence of Theorem 1.1 in \cite{JOP3}. Nevertheless, we provide a 
simple and more direct proof.

It clearly suffices to consider the case where $A$ and $B$ are monomials. Using the CAR, one can 
further restrict ourselves to $A$'s which are products of factors of the form $a^\ast(f_1)a(f_2)$. Finally,
using the identities $[AB,C]=A[B,C]+[A,C]B$ and $[A,B^\ast]=-[A^\ast,B]^\ast$, it suffices to consider 
the case $A=a^\ast(f_1)a(f_2)$ and $B=a^\ast(g)$. Since
$$
[a^\ast(f_1)a(f_2),\tau_{H_\bv}^t(a^\ast(g))]=a^\ast(f_1)\langle f_2|\e^{\i th_\bv}g\rangle,
$$
the result now follows from lemma \ref{LocDec}.\hfill$\square$

\subsection{Proof of Theorem \ref{pthm}}

\noindent{\bf Part (1).} By Remark 1 in Section \ref{presect} and Lemma \ref{LocDec} one has
$$
C_\bv=\sup_{t\in\rr,f,g\in\mathcal{D}_\bv}\left[
\frac{|\langle f|\e^{\i th_\bv}g\rangle|}{ \|\jx^3 f\|\,\|\jx^3 g\|}\langle t\rangle^{3/2}\right]<\infty.
$$
It follows from Theorem 1.2 of \cite{JOP3} that there exists a constant $C_W$, depending only on
the interaction $W$ such that for $|\xi|<\bar\xi_\bv=C_W/C_\bv$, the uniform limit
\begin{equation}
\varsigma_\bv(A)=\lim_{t\to\infty}\tau_{H_\bv}^{-t}\circ\tau_{K_\bv}^t(A),
\label{gammaW}
\end{equation}
exists for all $A\in\CAR(\fh)$. Moreover, $\varsigma_\bv$ is a $\ast$-automorphism 
of $\CAR(\fh)$ and
\begin{equation}
\varsigma_\bv^{-1}(A)=\lim_{t\to\infty}\tau_{K_\bv}^{-t}\circ\tau_{H_\bv}^{t}(A),
\label{gammaWinv}
\end{equation}
for all $A\in\CAR(\fh)$.

Since the range of the M\o ller operator $\omega_\bv$ is $\fh_\cR$, one has
$$
\slim_{t\to\infty}\e^{-\i th_{\cR,\bv}}\e^{\i th_\bv}=\omega_\bv.
$$
It follows from the uniform continuity of the map $\fh\ni f\mapsto a^\#(f)\in\CAR(\fh)$ that
$$
\lim_{t\to\infty}\tau_{\cR,\bv}^{-t}\circ\tau_{H_\bv}^t(a^\#(f))
=\lim_{t\to\infty}a^\#(\e^{-\i th_{\cR,\bv}}\e^{\i th_\bv}f)
=a^\#(\omega_\bv f)=\gamma_{\omega_\bv}(a^\#(f)).
$$
Thus, since the maps $\tau_{\cR,\bv}^{-t}\circ\tau_{H_\bv}^t$ and $\gamma_{\omega_\bv}$
are isometric $\ast$-morphisms,
\begin{equation}
\lim_{t\to\infty}\tau_{\cR,\bv}^{-t}\circ\tau_{H_\bv}^t(A)=\gamma_{\omega_\bv}(A),
\label{gammaomega}
\end{equation}
holds for any polynomial $A$ in the $a^\#$, and extends by density/continuity to all $A\in\CAR(\fh)$.

Combining \eqref{gammaW} and \eqref{gammaomega} and using again the isometric
nature of the various maps involved, we obtain
$$
\varsigma(A)=
\lim_{t\to\infty}\tau_{\cR,\bv}^{-t}\circ\tau_{K_\bv}^t(A)
=\lim_{t\to\infty}(\tau_{\cR,\bv}^{-t}\circ\tau_{H_\bv}^t)\circ(\tau_{H_\bv}^{-t}\circ\tau_{K_\bv}^t)(A)
=\gamma_{\omega_\bv}\circ\varsigma_\bv(A),
$$
for all $A\in\CAR(\fh)$. Since $\varsigma$ is the composition of two $\ast$-isomorphisms, it
is itself a $\ast$-isomorphism.

\bigskip
\noindent{\bf Part (2).}
Let $\st$ be an almost-$(\bbeta,\bmu)$-KMS state. 
For any $A\in\CAR(\fh)$ we can write
\beq
\langle\tau_{K_\bv}^t(A)\rangle=\langle\tau_{\cR,\bv}^t\circ
(\tau_{\cR,\bv}^{-t}\circ\tau_{K_\bv}^t)(A)\rangle,
\label{pstart}
\eeq
and Part (1) yields
$$
\lim_{t\to\infty}\left[\langle\tau_{K_\bv}^{t}(A)\rangle
-\langle\tau_{\cR,\bv}^{t}(\varsigma(A))\rangle\right]=0.
$$
Since the spectrum of $h_{\cR,\bv}$ acting on $\fh_\cR$ is purely absolutely continuous and the 
restriction of the state $\st_\cR^{\bbeta,\bmu}$ to $\CAR(\fh_\cR)$ is 
$\tau_{\cR,\bv}$-invariant, the $C^\ast$-dynamical system 
$(\CAR(\fh_\cR),\tau_{\cR,\bv},\st_\cR^{\bbeta,\bmu})$
is mixing (see e.g., \cite{AJPP1,Pi}). Using the facts that ${\rm Ran}(\varsigma)\subset\CAR(\fh_\cR)$ 
and that the restriction to this subalgebra of the initial state $\st$ is normal
w.r.t.\;the restriction of $\st_\cR^{\bbeta,\bmu}$ we conclude that
$$
\lim_{t\to\infty}\langle\tau_{K_\bv}^{t}(A)\rangle
=\lim_{t\to\infty}\langle\tau_{\cR,\bv}^{t}(\varsigma(A))\rangle
=\langle\varsigma(A)\rangle^{\bbeta,\bmu}_\cR.
$$

\noindent{\bf Part (3).} For $A\in{\mathcal A}_\bv$ we can invoke Theorem 1.1 in \cite{JOP3}
to conclude that the Dyson expansion \eqref{DysonStart}  converges uniformly in $t$ and
provides a convergent expansion of the map $\varsigma_\bv$. More precisely, one has
$$
\int_{\Delta_n}\left\|
\i[\tau_{H_\bv}^{-s_n}(W),\i[\tau_{H_\bv}^{-s_{n-1}}(W),\i[\cdots,\i[\tau_{H_\bv}^{-s_1}(W),A]\cdots]]]\right\|
\d s_1\cdots\d s_n<\infty,
$$
for all $n\ge1$ and the power series
\beq
\varsigma_\bv(A)=A+\sum_{n=1}^\infty\xi^n\int_{\Delta_n}
\i[\tau_{H_\bv}^{-s_n}(W),\i[\tau_{H_\bv}^{-s_{n-1}}(W),\i[\cdots,\i[\tau_{H_\bv}^{-s_1}(W),A]\cdots]]]
\d s_1\cdots\d s_n,
\label{sigmavdyson}
\eeq
converges in norm for $|\xi|<\bar\xi_\bv$. Using the fact that 
$\gamma_{\omega_\bv}\circ\tau_{H_\bv}^t=\tau_{\cR,\bv}^t\circ\gamma_{\omega_\bv}$ yields
the result.

\subsection{Proof of Theorem \ref{pfthm}}

\noindent{\bf Part (1).} We claim that
\begin{equation}
\lim_{t_0\to-\infty}\alpha_{\b0,t_0}^{0,t_0}\circ\tau_{H_\b0}^{-t_0}(B)=\varsigma_\b0^{-1}(B),
\label{Thirring}
\end{equation}
holds for all $B\in\CAR(\fh)$. Since $\alpha_{\b0,t_0}^{0,t_0}\circ\tau_{H_\b0}^{-t_0}$ and
$\varsigma_\b0^{-1}$ are $\ast$-automorphisms, it suffices to prove \eqref{Thirring}
for $B=a(f)$ with $f\in\mathcal{D}_\b0$. Duhamel's formula yields
\begin{align*}
\alpha_{\b0,t_0}^{0,t}\circ\tau_{H_\b0}^{-t}(B)
&=B-\i\int_0^{t}
\alpha_{\b0,t_0}^{0,s}\left([W-(1-\chi(s/t_0))H_{\rm T},\tau_{H_\b0}^{-s}(B)]\right)\d s,\\[4pt]
\tau_{K_\b0}^{t}\circ\tau_{H_\b0}^{-t}(B)
&=B-\i\int_0^{t}\tau_{K_\b0}^{s}\left([W,\tau_{H_\b0}^{-s}(B)]\right)\d s.
\end{align*}
Subtracting these two identities at $t=t_0<0$, we obtain the estimate
\begin{align*}
\left\|\alpha_{\b0,t_0}^{0,t_0}\circ\tau_{H_\b0}^{-t_0}(B)-\tau_{K_\b0}^{t_0}\circ\tau_{H_\b0}^{-t_0}(B)\right\|
\le&\int_{t_0}^0\left\|\left(\alpha_{\b0,t_0}^{0,s}-\tau_{K_\b0}^{s}\right)([W,\tau_{H_\b0}^{-s}(B)])\right\|\d s\\[4pt]
+&\int_{t_0}^0\left\|[H_{\rm T},\tau_{H_\b0}^{-s}(B)]\right\| \,(1-\chi(s/t_0))\d s.
\end{align*}
The integrand in the first term on right hand side of this expression is bounded by
$2\|[W,\tau_{H_\b0}^{-s}(B)]\|$ which is in $L^1(\rr,\d s)$ by Lemma \ref{asymptoticabelian}. 
Moreover, for fixed $s\in\rr$, it vanishes as $t_0\to-\infty$ by Lemma \ref{localautonomy}.
Hence, the dominated convergence theorem allows us to conclude that the first integral
vanishes as $t_0\to-\infty$. A similar argument applies to the second integral. Taking
Equ.\;\eqref{gammaWinv} into account concludes the proof of our claim.

Given Equ.\;\eqref{Thirring} we immediately obtain, for $A\in\CAR(\fh)$,
$$
\lim_{t_0\to-\infty}
\left\|\tau_{H_\b0}^{t_0}\circ\alpha_{\b0,t_0}^{t_0,0}(A)-\varsigma_\b0(A)\right\|
=\lim_{t_0\to-\infty}
\left\|A-\alpha_{\b0,t_0}^{0,t_0}\circ\tau_{H_\b0}^{-t_0}\circ\varsigma_\b0(A)\right\|
=\left\|A-\varsigma_\b0^{-1}\circ\varsigma_\b0(A)\right\|=0.
$$
From this point, the proof can proceed as for Theorem \ref{pthm}.

\bigskip
\noindent{\bf Part (2).} Due to the fact that $V_\cR=\d\Gamma(v_\cR)$ is not a local observable 
(and even not an element of $\CAR(\fh)$) the proof is more delicate than that of Part (1).

Denote by $\alpha_{\cR,\bv,t_0}^{s,t}$ the non-autonomous quasi-free dynamics generated by the
time-dependent one-particle Hamiltonian $h_\cR+\chi(t/t_0)v_\cR$. We first show that
\begin{equation}
\lim_{t_0\to-\infty}\left\|\alpha_{\bv,t_0}^{0,t_0}\circ\alpha_{\cR,\bv,t_0}^{t_0,0}(A)
-\tau_{K_\bv}^{t_0}\circ\tau_{\cR,\bv}^{-t_0}(A)\right\|=0,
\label{flu}
\end{equation}
holds for all $A\in\CAR(\fh_\cR)$. Since its suffices to prove this with $A=a(f)$ for a dense set of
$f\in\fh_\cR$, we can assume that $\jx^\sigma f\in\fh_\cR$ for some $\sigma>5/2$. Duhamel's formula
yields
\begin{equation}
\begin{split}
\alpha_{\bv,t_0}^{0,t_0}\circ\alpha_{\cR,\bv,t_0}^{t_0,0}(a(f))
-\tau_{K_\bv}^{t_0}\circ\tau_{\cR,\bv}^{-t_0}(a(f))
&=\int_{t_0}^0\tau_{K_\bv}^s\left(\i[H_{\rm T},(\tau_{\cR,\bv}^{-s}-\alpha_{\cR,\bv,t_0}^{s,0})(a(f))]
\right)\d s\\[4pt]
&+\int_{t_0}^0(\tau_{K_\bv}^{s}-\alpha_{\bv,t_0}^{0,s})\left(\i[H_{\rm T},\alpha_{\cR,\bv,t_0}^{s,0}(a(f))]
\right)\d s.
\end{split}
\label{deltalpha}
\end{equation}
Using the explicit formula
\begin{equation}
\alpha_{\cR,\bv,t_0}^{s,0}(a(f))
=a\left(\e^{\i\int_s^0\chi(u/t_0)v_\cR\d u}\e^{-\i sh_\cR}f\right),
\label{headache}
\end{equation}
the CAR and Lemma \ref{asymptoticabelian}, one derives the estimate
$$
\left\|[H_{\rm T},(\tau_{\cR,\bv}^{-s}-\alpha_{\cR,\bv,t_0}^{s,0})(a(f))]\right\|
\le C_\sigma\sum_{j=1}^m|d_j|\,\|\jx^\sigma f\|\,
|1-\e^{\i v_j\int_s^0(1-\chi(u/t_0))\d u}|\,\langle s\rangle^{-3/2}.
$$
It follows from the dominated convergence theorem that the first integral on the right hand side of 
Equ.\;\eqref{deltalpha} vanishes as $t_0\to-\infty$. Due to Lemma \ref{localautonomy} and the estimate
$$
\left\|[H_{\rm T},\alpha_{\cR,\bv,t_0}^{s,0}(a(f))]\right\|
\le C_\sigma\sum_{j=1}^m|d_j|\,\|\jx^\sigma f\|\,\langle s\rangle^{-3/2},
$$
the same is true for the second integral and this proves \eqref{flu}.

Since the range of $\gamma_{\omega_\bv}$ is $\CAR(\fh_\cR)$, we have
\begin{align*}
\lim_{t_0\to-\infty}\left\|
\alpha_{\cR,\bv,t_0}^{0,t_0}\circ\alpha_{\bv,t_0}^{t_0,0}(A)
-\gamma_{\omega_\bv}\circ\varsigma_\bv(A)\right\|
&=\lim_{t_0\to-\infty}\left\|A-\alpha_{\bv,t_0}^{0,t_0}\circ\alpha_{\cR,\bv,t_0}^{t_0,0}
(\gamma_{\omega_\bv}\circ\varsigma_\bv(A))\right\|\\
&=\lim_{t_0\to-\infty}\left\|A-\tau_{K_\bv}^{t_0}\circ\tau_{\cR,\bv}^{-t_0}
(\gamma_{\omega_\bv}\circ\varsigma_\bv(A))\right\|\\
&=\lim_{t_0\to-\infty}\left\|\tau_{\cR,\bv}^{t_0}\circ\tau_{K_\bv}^{-t_0}(A)-
\gamma_{\omega_\bv}\circ\varsigma_\bv(A)\right\|\\
&=0,
\end{align*}
for any $A\in\CAR(\fh)$. Now it follows from \eqref{headache} that 
$\alpha_{\cR,\bv,t_0}^{0,t_0}=\tau_{\cR,\tilde\bv}^{t_0}$ with
$$
\tilde\bv=\bv \int_0^1\chi(s)\d s,
$$
so that we can write the following analogue of Equ.\;\eqref{pstart}
$$
\langle\alpha_{\bv,t_0}^{t_0,0}(A)\rangle
=\langle\tau_{\cR,\tilde\bv}^{-t_0}\circ
(\alpha_{\cR,\bv,t_0}^{0,t_0}\circ\alpha_{\bv,t_0}^{t_0,0})(A)\rangle,
$$
and finish the proof as for Theorem \ref{pthm}.

\subsection{Proof of Theorem \ref{entropicthm}}

The main arguments used in this section are simple adaptations of Section 3.6 and 3.7 in \cite{JOP2}.

\noindent{\bf Part (1).} 
We start with some basic facts from modular theory (the reader is referred to \cite{BR1} for a detailed
exposition). A state on $\CAR(\fh)$ is called modular
if it is a $(\beta,\mu)$-KMS state with $\beta=-1$ and $\mu=0$ for some strongly continuous group
$\sigma$ of $\ast$-automorphisms of $\CAR(\fh)$\footnote{The choice of $\beta=-1$ and $\mu=0$
is conventional in the mathematical literature and has no physical meaning}.
The state $\st_\cR^{\bbeta,\bmu}$
is modular and its modular group $\sigma$ is easily seen to be the quasi-free dynamics generated
by the one-particle Hamiltonian 
$$
k=-\sum_{j=1}^m\beta_j(h_j-\mu_j1_j).
$$
We denote by $\delta$ the generator of $\sigma$, i.e., $\sigma^t=\e^{t\delta}$.

For a self-adjoint $P\in\CARv(\fh)$ define the group $\sigma_P$ by
$\sigma_P^t(A)=\e^{\i t(\d\Gamma(k)+P)}A\e^{-\i t(\d\Gamma(k)+P)}$. By Araki's perturbation
theory $\sigma_P$ has a unique $(-1,0)$-KMS state which we denote $\st_P$.  
Let $\frak S$ be the set of all  states obtained in this way. This set is norm dense in the
set of $\st_\cR^{\bbeta,\bmu}$-normal states on $\CARv(\fh)$ 
(this is a consequence of the final remark in Section 5 of \cite{A1}). 

The fact that ${\rm S}(\st_P|\st_P)=0$ and the
fundamental formula of Araki (Theorem 3.10 in \cite{A2} or Proposition 6.2.32 in \cite{BR2}) yield
\beq
{\rm S}(\langle\tau_{K_\bv}^t(\,\cdot\,)\rangle_P|\st_P)
={\rm S}(\langle\tau_{K_\bv}^t(\,\cdot\,)\rangle_P|\st_{\cR}^{\beta,\mu})
-{\rm S}(\st_P|\st_{\cR}^{\beta,\mu})
+\langle\tau_{K_\bv}^t(P)-P\rangle_P.
\label{ArakiForm}
\eeq
Let $Q=H_\cS+\xi W+H_{\rm T}\in\CARv(\fh_\cS)$ and note that $K_\bv=H_{\cR,\bv}+Q$.
A simple calculation shows that $Q\in{\rm Dom}(\delta)$ and
\beq
\delta(Q)=\i[\d\Gamma(k),Q]=
\i[-\sum_{j=1}^m\beta_j(H_j-\mu_jN_j),K_v-H_{\cR,\bv}]
=-\sum_{j=1}^m\beta_j(E_j-\mu_jJ_j).
\label{deltaQ}
\eeq
Since $\st_{\cR}^{\beta,\mu}$ is $\tau_{\cR,\bv}$-invariant, we can
apply the entropy balance formula of \cite{JP3,JP4} to obtain
$$
{\rm S}(\langle\tau_{K_\bv}^t(\,\cdot\,)\rangle_P|\st_{\cR}^{\beta,\mu})=
{\rm S}(\st_P|\st_{\cR}^{\beta,\mu})
-\int_0^t\langle\tau_{K_\bv}^s(\delta(Q))\rangle_P\,\d s.
$$
Inserting this relation into \eqref{ArakiForm} further gives
\beq
{\rm S}(\langle\tau_{K_v}^{-t_0}(\,\cdot\,)\rangle_P|\st_P)=
\langle\tau_{K_v}^{-t_0}(P)-P\rangle_P
-\int_0^{-t_0}\langle\tau_{K_v}^s(\delta(Q))\rangle_P\,\d s.
\label{Ebal}
\eeq
Dividing by $t_0$ and taking the limit $t_0\to-\infty$ we get
$$
\lim_{t_0\to-\infty}\frac1{t_0}
{\rm S}(\langle\tau_{K_v}^{-t_0}(\,\cdot\,)\rangle_P|\st_P)
=\langle\delta(Q)\rangle_+^{\bbeta,\bmu,\bv},
$$
which, taking into account \eqref{deltaQ} and the fact that relative entropies
are non-positive, yields the result.

\bigskip
\noindent{\bf Part (2).} This is just a special case of Part (1). It suffices to notice that
if $\bbeta=(\beta,\ldots,\beta)$ and $\bmu=(\mu,\ldots,\mu)$ then $k=-\beta(h_\cR-\mu)$
and hence $\d\Gamma(k)+P=-\beta(K_\b0-\mu N)$ where $P=-\beta(Q-\mu N_\cS)\in\CARv(\fh)$.
It follows that $\st_{K_\b0}^{\beta,\mu}=\st_P\in{\frak S}$.
Finally, taking into account the sum rules \eqref{SumRules} we get
$$
-\sum_{j=1}^m\beta\langle E_j-\mu J_j\rangle_+^{\bbeta,\bmu,\bv}
=\sum_{j=1}^m\beta v_j\langle J_j\rangle_+^{\bbeta,\bmu,\bv}.
$$
We note for later reference that Equ.\;\eqref{Ebal} yields the following entropy balance relation for
the partition-free NESS
\beq
{\rm S}(\langle\tau_{K_v}^{t}(\,\cdot\,)\rangle_{K_\b0}^{\beta,\mu}|\st_{K_\b0}^{\beta,\mu})=
\langle\tau_{K_v}^{t}(P)-P\rangle_{K_\b0}^{\beta,\mu}
-\int_0^{t}\langle\tau_{K_v}^s(\delta(Q))\rangle_{K_\b0}^{\beta,\mu}\,\d s.
\label{PFbal}
\eeq

\bigskip
\noindent{\bf Part (3).} Let $\bbeta=(\beta,\ldots,\beta)$, $\bmu=(\mu,\ldots,\mu)$ and
$\bv_0=(v_0,\ldots,v_0)$ and set $\widetilde{K}_\b0=K_\b0-v_0N_\cS$.
It follows from the identity
$$
K_{\bv_0}=K_\b0+v_0\sum_{j=1}^mN_j=\widetilde{K}_\b0+v_0N,
$$
that $\tau_{K_{\bv_0}}^t=\tau_{\widetilde{K}_\b0}^t\circ\vartheta^{tv_0}$. The gauge invariance 
$\vartheta^s\circ\tau_{\cR,\bv_0}^{-t}\circ\tau_{K_{\bv_0}}^t
=\tau_{\cR,\bv_0}^{-t}\circ\tau_{K_{\bv_0}}^t\circ\vartheta^s$
and the fact that $\varsigma$ maps onto $\CAR(\fh_\cR)$ implies
$\varsigma\circ\vartheta^t=\vartheta_\cR^t\circ\varsigma$ where $\vartheta_\cR$ is the gauge
group of $\CAR(\fh_\cR)$ (i.e., the quasi-free dynamics on $\CAR(\fh_\cR)$ generated by
$1_\cR$). Together with the intertwining property of $\varsigma$,  this yields 
$$
\varsigma\circ(\tau_{\widetilde{K}_\b0}^t\circ\vartheta^{-t\mu})
=\varsigma\circ\tau_{K_{\bv_0}}\circ\vartheta^{-t(\mu+v_0)}=
\tau_{\cR,\bv_0}^t\circ\varsigma\circ\vartheta^{-t(\mu+v_0)}
=(\tau_{\cR}^t\circ\vartheta_\cR^{-t\mu})\circ\varsigma.
$$
It follows that for any $A,B\in\CAR(\fh)$ one has
$$
\langle A\,\tau_{\widetilde{K}_\b0}^t\circ\vartheta^{-t\mu}(B)\rangle_+^{\bbeta,\bmu,\bv_0}
=\langle \varsigma(A)\varsigma(\tau_{\widetilde{K}_\b0}^t\circ\vartheta^{-t\mu}(B))\rangle_\cR^{\bbeta,\bmu}
=\langle \varsigma(A)\tau_\cR^t\circ\vartheta_\cR^{-t\mu }(\varsigma(B))\rangle_\cR^{\bbeta,\bmu},
$$
and since $\st_\cR^{\bbeta,\bmu}$ is $(\beta,\mu)$-KMS for $\tau_\cR$
one easily concludes that $\st_+^{\bbeta,\bmu,\bv_0}$ satisfies the
$(\beta,\mu)$-KMS condition for $\tau_{\widetilde{K}_\b0}$. Since the later group is a local perturbation
of the quasi-free dynamics $\tau_{H_\b0}$, it follows from Araki's perturbation theory that the 
partition-free NESS is the unique $(\beta,\mu)$-KMS state for $\tau_{\widetilde{K}_\b0}$ which we
denote $\st_{\widetilde{K}_\b0}^{\beta,\mu}$.

Observe that replacing the one-particle Hamiltonian $h_\cS$ by $h_\cS-v_0 1_\cS$
transforms $K_\b0$ into $\widetilde{K}_\b0$. The same substitution changes $K_\bv$ into
$\widetilde{K}_\bv=K_{\bv_0+\bv}-v_0N=K_v-v_0N_\cS$ while the entropy balance relation 
\eqref{PFbal} transforms into
$$
{\rm S}(\langle\tau_{\widetilde{K}_\bv}^t(\,\cdot\,)\rangle_{\widetilde{K}_\b0}^{\beta,\mu}
|\st_{\widetilde{K}_\b0}^{\beta,\mu})
=\langle\tau_{\widetilde{K}_\bv}^t(\widetilde{P})-\widetilde{P}\rangle_{\widetilde{K}_\b0}^{\beta,\mu}
-\int_0^t\langle\tau_{\widetilde{K}_\bv}^s(\delta(\widetilde{Q}))\rangle_{\widetilde{K}_\b0}^{\beta,\mu}\,\d s,
$$
where $\widetilde{Q}=Q-v_0N_\cS$ and $\widetilde{P}=P+\beta v_0N_\cS$. Dividing this relation
by $t>0$ and letting $t\downarrow0$ we obtain, after some elementary algebra and using the
fact that relative entropies are non-positive
$$
0\ge
\lim_{t\downarrow0}\frac1t{\rm S}(\langle\tau_{\widetilde{K}_\bv}^t(\,\cdot\,)\rangle_{\widetilde{K}_\b0}^{\beta,\mu}|
\st_{\widetilde{K}_\b0}^{\beta,\mu})
=\langle\i[\widetilde{K}_\bv,\widetilde{P}]-\delta(\widetilde{Q})\rangle_{\widetilde{K}_\b0}^{\beta,\mu}
=-\beta\sum_{j=1}^m v_j\langle J_j\rangle_{\widetilde{K}_\b0}^{\beta,\mu}.
$$
Since this relation holds for all $\bv\in\rr^m$  and $\langle J_j\rangle_{\widetilde{K}_\b0}^{\beta,\mu}$
does not depend on $\bv$ we can conclude that
$\langle J_j\rangle_{\widetilde{K}_\b0}^{\beta,\mu}=0$ for all $j$.

To deal with the energy currents, we set 
$K^\natural_{\boldsymbol{\alpha}}=\widetilde{K}_\b0+\sum_{j=1}^m\alpha_j H_j$ with
$\boldsymbol{\alpha}=(\alpha_1,\ldots,\alpha_m)\in\rr^m$ and invoke the same arguments to
derive the inequality
$$
0\ge
\lim_{t\downarrow0}\frac1t{\rm S}(\langle\tau_{K_{\boldsymbol{\alpha}}^\natural}^t(\,\cdot\,)\rangle_{\widetilde{K}_\b0}^{\beta,\mu}|
\st_{\widetilde{K}_\b0}^{\beta,\mu})
=\langle\i[K_{\boldsymbol{\alpha}}^\natural,\widetilde{P}]-\delta(\widetilde{Q})\rangle_{\widetilde{K}_\b0}^{\beta,\mu}
=-\beta\sum_{j=1}^m\alpha_j\langle E_j\rangle_{\widetilde{K}_\b0}^{\beta,\mu},
$$
from which we conclude that $\langle E_j\rangle_{\widetilde{K}_\b0}^{\beta,\mu}=0$ for all $j$.


\subsection{Proof of Theorem \ref{specthm}}

We only consider the lesser Green-Keldysh function. The case of the greater function
is completely similar.

\noindent{\bf Parts (1)--(3).}
We observe that $A_x=\varsigma(a_x)$ where the M\o ller morphism $\varsigma$ is given
by $\varsigma=\gamma_{\omega_\bv}\circ\varsigma_\bv$. Since the
range of $\gamma_{\omega_\bv}$ is $\CAR(\fh_\cR)$ one has $A_x\in\CAR(\fh_\cR)$.
Using the intertwining property of $\varsigma$, Theorem \ref{pthm} allows us to write
$$
G^{<\bbeta,\bmu,\bv}_+(t;x,y)=\i\langle a^\ast_y\tau_{K_\bv}^t(a_x)\rangle_+^{\bbeta,\bmu,\bv}
=\i\langle\varsigma(a^\ast_y\tau_{K_\bv}^t(a_x))\rangle_\cR^{\bbeta,\bmu}
=\i\langle\varsigma(a^\ast_y)\tau^t_{\cR,\bv}(\varsigma(a_x))\rangle_\cR^{\bbeta,\bmu}
=\i\langle A_y^\ast\tau^t_{\cR,\bv}(A_x)\rangle_\cR^{\bbeta,\bmu}.
$$
Passing to the GNS representation and using the fact that $L_{\cR,\bv}\Omega_\cR=0$ we obtain
$$
G^{<\bbeta,\bmu,\bv}_+(t;x,y)
=\i(\Omega_\cR|\pi_\cR(A_y)^\ast\e^{\i tL_{\cR,\bv}}\pi_\cR(A_x)\e^{-\i tL_{\cR,\bv}}\Omega_\cR)
=\i(\Psi_y|\e^{\i tL_{\cR,\bv}}\Psi_x)
=\i\int_\rr\e^{\i t\omega}\d\lambda_{\Psi_y,\Psi_x}(\omega),
$$
where $\Psi_x=\pi_\cR(A_x)\Omega_\cR$ and $\lambda_{\Psi_y,\Psi_x}$ denotes the
spectral measure of $L_{\cR,\bv}$ for $\Psi_y$ and $\Psi_x$. We note that
$$
(\Omega_\cR|\Psi_x)=(\Omega_\cR|\pi_\cR(A_x)\Omega_\cR)
=\langle A_x\rangle_\cR^{\bbeta,\bmu}=\langle a_x\rangle_+^{\bbeta,\bmu,\bv}=0,
$$
since the NESS $\st_+^{\bbeta,\bmu,\bv}$ is gauge-invariant. This proves Part (1). Moreover, this
implies that the spectral measure $\lambda_{\Psi_x,\Psi_y}$ is absolutely continuous 
w.r.t.\;Lebesgue's measure so that
$$
G^{<\bbeta,\bmu,\bv}_+(t;x,y)=\i\int_\rr\e^{\i t\omega}
\frac{\d\lambda_{\Psi_y,\Psi_x}(\omega)}{\d\omega}\,\d\omega,
$$
which proves Part (2).

\bigskip
\noindent{\bf Part (4).}
We first note that $a_x$ is an entire analytic element for the group $\tau_{K_\bv}$.
This is an simple consequence of the interaction picture representation
$$
\tau_{K_\bv}^t(a_x)=\Gamma_\bv^t\tau_{H_\bv}^t(a_x)\Gamma_\bv^{t\ast},
$$
where the cocycle $\Gamma_\bv^t=\e^{\i tK_\bv}\e^{-\i tH_\bv}$ has the Dyson expansion
$$
\Gamma_\bv^t=I+\sum_{n=1}^\infty(\i\xi t)^n \int\limits_{0\le s_1\le\cdots\le s_n\le 1}
\tau_{H_\bv}^{ts_1}(W)
\cdots\tau_{H_\bv}^{ts_n}(W)\d s_1\cdots\d s_n.
$$
Indeed, since $h_\bv$ is bounded, $\tau_{H_\bv}^t(a_x)=a(\e^{\i th_\bv}\delta_x)$
and $\tau_{H_\bv}^t(W)$ extend to entire analytic functions of $t$ and the above Dyson expansion
converge in norm for any complex value of $t$, defining an entire analytic $\CAR(\fh)$-valued function.
Finally, since
$$
\partial_z^n G^{<\bbeta,\bmu,\bv}_+(z+t;x,y)
=\i\langle a_y^\ast\tau_{K_\bv}^t(A_z)\rangle_+^{\bbeta,\bmu,\bv},
$$
where $A_z=\partial_z^n\tau_{K_\bv}^z(a_x)\in\CAR(\fh)$, the last assertion follows from the
mixing property of the dynamical system $(\CAR(\fh),\tau_{K_\bv},\st_+^{\bbeta,\bmu,\bv})$
(Remark 4, Section \ref{sct-part}).

\bigskip\noindent{\bf Parts (5)--(6).}
By Theorem 1.1 of \cite{JOP3} the Dyson expansion~\eqref{sigmavdyson} can be reorganized as
$$
\varsigma_\bv(a_x)=\sum_{n=0}^\infty\xi^n\sum_{q\in\mathcal{Q}_n}\int_{\Delta_n}
G^{(n)}_{x,q}(s)F^{(n)}_{q}(s)\d s,
$$
where the $\mathcal{Q}_n$ are finite sets, the $G^{(n)}_{x,q}(s)$ scalar functions of 
$s=(s_1,\ldots,s_n)\in\rr^n$ such that
\beq
\sum_{n=0}^\infty\bar\xi_\bv^n\sum_{q\in\mathcal{Q}_n}\int_{\Delta_n}|G^{(n)}_{x,q}(s)|\d s<\infty,
\label{finitesum}
\eeq
and the $F^{(n)}_{q}(s)$ are monomials of degree $k_q^{(n)}$
with factors in $\{a^\#(\e^{-\i uh_\bv}\delta_z)\,|\,z\in\cS,u\in\{s_1,\ldots,s_n\}\}$. Moreover,
the $k_q^{(n)}$ are odd and satisfy $k_q^{(n)}\le nk_W+1$ for some integer $k_W$
depending only on the interaction $W$.

Setting $\st_+=\st_{+,\xi=0}^{\bbeta,\bmu,\bv}$, 
the identity $\langle a_y^\ast \tau_{K_\bv}^t(a_x)\rangle_+^{\bbeta,\bmu,\bv}
=\langle\varsigma_\bv(a_y^\ast)\tau_{H_\bv}^t(\varsigma_\bv(a_x))\rangle_+$ leads to the
estimate
\beq
\begin{split}
\int_{-\infty}^\infty\left|G_+^{<\bbeta,\bmu,\bv}(t+\i\eta,x,y)\right|\ dt
\le\sum_{n_1,n_2=0}^\infty&|\xi|^{n_1+n_2}
\sum_{q_1\in\mathcal{Q}_{n_1}\atop q_2\in\mathcal{Q}_{n_2}}
\left[\int\limits_{\Delta_{n_1}} |G^{(n_1)}_{y,q_1}(s)|\d s\right]
\left[\int\limits_{\Delta_{n_2}} |G^{(n_2)}_{x,q_2}(s')|\d s'\right]\\
&\times\,\sup_{(s,s')\in\rr^{n_1}\times\rr^{n_2}}\int_{-\infty}^\infty
|\langle F^{(n_1)}_{q_1}(s)^\ast\tau_{H_\bv}^{t+\i\eta}(F^{(n_2)}_{q_2}(s'))\rangle_+|\d t.
\end{split}
\label{bigintform}
\eeq
To estimate the last integral we denote by $k_j$ the degree of the monomial $F_{q_j}^{(n_j)}$ 
and expand each factor of this monomial in terms of field operators using the identity
$a^\#(f)=2^{-1/2}(\varphi(f)\pm\i\varphi(\i f))$. In this way we can
write $\langle F^{(n_1)}_{q_1}(s)^\ast\tau_{H_\bv}^{t+\i\eta}(F^{(n_2)}_{q_2}(s'))\rangle_+$ 
as a sum of $2^{k_1+k_2}$ terms of the form
$2^{-(k_1+k_2)/2}\langle\varphi(f_1)\cdots\varphi(f_{k_1+k_2})\rangle_+$,
where
$$
f_i\in\left\{
\begin{array}{lcl}
\{\e^{-\i uh_\bv}\delta_z\,|\,z\in\cS,u\in\{s_1,\ldots,s_{n_1}\}\}&\text{ if }
&i\in \mathcal{V}_1=\{1,\ldots,k_1\},\\[6pt]
\{\e^{\i(t-u'\pm\i\eta)h_\bv}\delta_z\,|\,z\in\cS,u'\in\{s_1',\ldots,s_{n_2}'\}\}&\text{ if }
&i\in \mathcal{V}_2=\{k_1+1,\ldots,k_1+k_2\}.
\end{array}
\right.
$$
Since $k_1$ and $k_2$ are odd, each term in the Wick expansion \eqref{wickform} of
$\langle\varphi(f_1)\cdots\varphi(f_{k_1+k_2})\rangle_+$ contains a factor 
$\langle\varphi(f_i)\varphi(f_j)\rangle_+$ such that $i\in\mathcal{V}_1$
and $j\in\mathcal{V}_2$. For such a pair $(i,j)$
denote by $\mathcal{P}_{i,j}$ the set of pairings of elements of 
$(\mathcal{V}_1\cup\mathcal{V}_2)\setminus\{i,j\}$
and for $p\in\mathcal{P}_{i,j}$ let $p\vee(i,j)$ be the pairing on $\mathcal{V}_1\cup\mathcal{V}_2$
obtained by merging $p$ with the pair $(i,j)$. Invoking Lemma 4.1 of \cite{JPP} we can write
\begin{align*}
\langle\varphi(f_1)\cdots\varphi(f_{k_1+k_2})\rangle_+
&=\sum_{i\in\mathcal{V}_1}\sum_{j\in\mathcal{V}_2}\langle\varphi(f_i)\varphi(f_j)\rangle_+
\sum_{p\in\mathcal{P}_{i,j}}\varepsilon(p\vee(i,j))
\prod_{(k,l)\in p}\langle\varphi(f_k)\varphi(f_l)\rangle_+\\
&=\sum_{i\in\mathcal{V}_1}\sum_{j\in\mathcal{V}_2}\varepsilon_{i,j}\langle\varphi(f_i)\varphi(f_j)\rangle_+
\sum_{p\in\mathcal{P}_{i,j}}\varepsilon(p)
\prod_{(k,l)\in p}\langle\varphi(f_k)\varphi(f_l)\rangle_+\\
&=\sum_{i\in\mathcal{V}_1}\sum_{j\in\mathcal{V}_2}\varepsilon_{i,j}\langle\varphi(f_i)\varphi(f_j)\rangle_+
\langle\varphi(f_1)\cdots\bcancel{\varphi(f_i)}\cdots\bcancel{\varphi(f_j)}\cdots\varphi(f_{k_1+k_2})\rangle_+,
\end{align*}
where $|\varepsilon_{i,j}|=1$. Using the facts  that $\|\varphi(f)\|=2^{-1/2}\|f\|$,
$\langle\varphi(f_i)\varphi(f_j)\rangle_\varrho=2^{-1}(\langle f_i|(I-\varrho)f_j\rangle
+\langle f_j|\varrho f_i\rangle)$ and Lemma \ref{LocDec} we get the estimate
$$
|\langle\varphi(f_i)\varphi(f_j)\rangle_+
\langle\varphi(f_1)\cdots\bcancel{\varphi(f_i)}\dots\bcancel{\varphi(f_j)}\cdots\varphi(f_{k_1+k_2})\rangle_+|
\le 
C\e^{c|\eta|k_2}2^{-(k_1+k_2)/2}\langle t-u\rangle^{-3/2}
$$
for some constants $C$ and $c$ and some $u\in\rr$. Integrating over $t$ before summing all 
contributions yields, taking into account the bounds $k_j\le n_jk_W+1$,
$$
\int_{-\infty}^\infty
|\langle F^{(n_1)}_{q_1}(s)^\ast\tau_{H_\bv}^{t+\i\eta}(F^{(n_2)}_{q_2}(s'))\rangle_+|\d t
\le C' \e^{c'|\eta|n_2} n_1n_2,
$$
for some constants $C'$ and $c'$. The estimate \eqref{finitesum} allows us to conclude that the 
integral on the left-hand side of \eqref{bigintform} is finite provided $|\xi|\e^{c'|\eta|}<\bar\xi_\bv$ 
so that Part (5) holds for any $\theta<\log(\bar\xi_\bv/|\xi|)/c'$. 
Part (6) follows from a Paley-Wiener argument (see, e.g., Theorem IX.14 in \cite{RS2}).

\subsection{Proof of Theorem \ref{HF-thm}}

\noindent{\bf Part (1).} 
Note that the function $\rr\ni E\mapsto v_\cS(E,\bv)$ is continuous and vanishes at infinity. Thus,
if Assumption \SPv{} is satisfied then $C_\cS=\sup_{E\in\rr}\|{\mathfrak m}_\bv(E)\|<\infty$. Since the
matrix
$$
{\mathfrak m}_{{\rm HF},\bv}(E)=(h_\cS+\xi v_{\rm HF}+v_\cS(E,\bv)-E)^{-1},
$$
exists for all $E\in\rr$ if $|\xi|<(C_\cS\|v_{\rm HF}\|)^{-1}$, Part (1) follows from 
Theorem \ref{pthm}.

\bigskip
\noindent{\bf Part (2).}  To simplify notation let $\st_+$ denote the non-interacting NESS
$\langle\,\cdot\,\rangle_{+,\xi=0}^{\bbeta,\bmu,\bv}$. Starting with the fact that
$\langle A\tau_{K_\bv}^t(B)\rangle_+^{\bbeta,\bmu,\bv}
=\langle\varsigma_\bv(A)\tau_{H_\bv}^t(\varsigma_\bv(B))\rangle_{+}$ and using the
$\tau_{H_\bv}$-invariance of $\st_+$, the Dyson expansion \eqref{sigmavdyson} yields
$$
\langle A\tau_{K_\bv}^t(B)\rangle_{+}^{\bbeta,\bmu,\bv}
=\langle A\tau_{H_\bv}^t(B)\rangle_+
+\i\xi\int_0^\infty\langle [W,\tau_{H_\bv}^{s}(A)]\tau_{H_\bv}^{s+t}(B)
+\tau_{H_\bv}^{s-t}(A)[W,\tau_{H_\bv}^s(B)]\rangle_+\,\d s+{\mathcal O}(\xi^2),
$$
for $A,B\in{\mathcal A}_\bv$. In the same way, we get
$$
\langle A\tau_{K_\bv}^t(B)\rangle_{\rm HF+}^{\bbeta,\bmu,\bv}
=\langle A\tau_{H_\bv}^t(B)\rangle_+
+\i\xi\int_0^\infty\langle [W_{\rm HF},\tau_{H_\bv}^{s}(A)]\tau_{H_\bv}^{s+t}(B)
+\tau_{H_\bv}^{s-t}(A)[W_{\rm HF},\tau_{H_\bv}^s(B)]\rangle_+\,\d s+{\mathcal O}(\xi^2).
$$
Since the dynamical system $(\CAR(\fh),\tau_{H_\bv},\st_+)$ is quasi-free and
$\langle A\,\i[W,B]\rangle_+=\overline{\langle A^\ast\,\i[W,B^\ast]\rangle_+}$ holds for
any  $A,B,W\in\CAR(\fh)$ with $W=W^\ast$, the proof of the estimate \eqref{HFform}
reduces to showing that
\beq
\langle[W-W_{\rm HF},A]B\rangle_\varrho=0,
\label{HFmagic}
\eeq
holds with $A=a(f)$, $B=a^\ast(g)$ and $A=a^\ast(f)$, $B=a(g)$ for any $f,g\in\fh$ and any density
operator $\varrho$. Now a simple calculation yields
\begin{align*}
\langle[W,a^\ast(f)]a(g)\rangle_\varrho&=\langle g|\varrho v_{\rm HF}f\rangle
=\langle[W_{\rm HF},a^\ast(f)]a(g)\rangle_\varrho\\
\langle[W,a(g)]a^\ast(f)\rangle_\varrho&=-\langle g|v_{\rm HF}(I-\varrho)f\rangle
=\langle[W_{\rm HF},a(g)]a^\ast(f)\rangle_\varrho,
\end{align*}
thus establishing the validity of Relation \eqref{HFmagic}. The local uniformity of the error term
in Equ.\;\eqref{HFform} easily follows from the estimate (1.5) in \cite{JOP3}.

\section{Conclusions and open problems}
\label{sectiunea6}

To the best of our knowledge, we provide for the first time sufficient conditions ensuring 
the existence of a steady state regime for the Green-Keldysh correlation functions of interacting 
fermions in mesoscopic systems in the partitioning and partition-free scenarios. Our proof
handles these two cases in a unified way and even allows for mixed, thermodynamical and
mechanical drive as well as adiabatic switching of these drives. We also show that the
steady state, when it exists, is largely insensitive of the initial state of the system, depending
only on its gross thermodynamical properties and not on structural properties like being
a product state or a quasi-free state.

Roughly speaking, the most important technical conditions which insure the existence of a
steady-state are two: {\it (i)} the non-interacting but fully coupled model has no bound states, 
and {\it (ii)} the strength of the self-interaction is small enough. 

Under these conditions we do not have to perform an ergodic limit.  As a practical 
application, we have shown that, up to second order corrections in the interaction strength, 
steady charge and energy currents coincide with their Hartree-Fock approximation which can 
be expressed in terms of a Landauer-B\"uttiker formula.

Let us point out some future steps towards a complete mathematical 
formulation of interacting quantum transport. Perhaps the most important 
progress would be to extend our scattering formalism to systems with bound states 
(i.e. weakly coupled quantum dots in the Coulomb blockade regime) and to a 
give a rigorous account on the diagrammatic recipes for the Green-Keldysh functions 
and interaction self-energies.
 
Another challenging issue is the existence of NESS for strongly correlated systems. 
This regime leads to the well known mesoscopic Kondo effect which relies on the Coulomb interaction 
between localized spins on the dot and the incident electrons from the leads. The theoretical 
treatment of this effect is notoriously difficult as the underlying Kondo and Anderson Hamiltonians 
do not allow perturbative calculations with respect to the interaction strength (see, e.g., the 
reviews \cite{PG,H} and \cite{KAO}).


\begin{thebibliography}{XXXXX} 

\bibitem[A1]{A1} Araki, H.:
Relative Hamiltonian for faithful normal states of a von Neumann algebra.
Publ. RIMS, Kyoto Univ. {\bf 9}, 165--209 (1973).

\bibitem[A2]{A2} Araki, H.:
Relative entropy for states of von Neumann algebras II.
Publ. RIMS, Kyoto Univ. {\bf 13}, 173--192 (1977).

\bibitem[ABGK]{ABGK} Avron, J.E., Bachmann, S., Graf, G.M. and Klich, I.:
Fredholm determinants and the statistics of charge transport. 
Commun. Math. Phys. {\bf 280}, 807--829 (2008).

\bibitem[AH]{AH} Araki, H., and Ho, T.G.:
Asymptotic time evolution of a partitioned infinite two-sided isotropic XY-chain.
Proc. Steklov Inst. Math. {\bf 228},191--204 (2000).

\bibitem[AJPP1]{AJPP1} Aschbacher, W., Jak\v{s}i\'c, V., Pautrat, Y., and Pillet, C.-A:
Topics in non-equilibrium quantum statistical mechanics.
In {\em Open Quantum Systems III. Recent Developments.}
S. Attal, A. Joye and C.-A. Pillet editors. Lecture Notes in Mathematics {\bf 1882}. 
Springer, Berlin, 2006.

\bibitem[AJPP2]{AJPP2} Aschbacher, W., Jak\v{s}i\'c, V., Pautrat, Y., and Pillet, C.-A.:
Transport properties of quasi-free Fermions.
J. Math. Phys. {\bf 48}, 032101-1--28 (2007).

\bibitem[AP]{AP} Aschbacher, W., and Pillet, C.-A.: 
Non-equilibrium steady states of the XY chain.
J. Stat. Phys. {\bf 112}, 1153--1175 (2003).

\bibitem[AS]{AS} Aschbacher, W., and Spohn, H: 
A remark on the strict positivity of entropy production.
Lett. Math. Phys. {\bf 75}, 17--23 (2006).

\bibitem[AW]{AW} Araki, H., and Wyss, W.:
Representations of canonical anticommutation relations. 
Helv. Phys. Acta {\bf 37}, 139--159 (1964).

\bibitem[BMa]{BMa} Botvich, D.D, and Maassen, H.:
A Galton--Watson estimate for Dyson series.
Ann. Henri Poincar\'e {\bf 10}, 1141--1158 (2009).

\bibitem[BM]{BM} Botvich, D.D., and Malyshev, V.A.: 
Unitary equivalence of temperature dynamics for ideal and locally perturbed Fermi gas.
Commun. Math. Phys. {\bf 91}, 301-- 312 (1983).

\bibitem[BR1]{BR1} Bratelli, O., and Robinson, D.W.:
{\it Operator Algebras and Quantum Statistical Mechanics 1.} 
Second Edition. Springer, New York,1997.

\bibitem[BR2]{BR2} Bratelli, O., and Robinson, D.W.:
{\it Operator Algebras and Quantum Statistical Mechanics 2.} 
Second Edition. Springer, New York,1997.

\bibitem[CCNS]{CCNS} Caroli, C., Combescot, R., Nozi\`eres, P., and Saint-James, D.:
Direct calculation of the tunneling current.
J. Phys. C: Solid State Phys. {\bf 4}, 916 (1971).

\bibitem[Ci]{Ci} Cini, M.:
Time-dependent approach to electron transport through junctions: General 
theory and simple applications. 
Phys. Rev. B. {\bf 22}, 5887 (1980).

\bibitem[CDNP]{CDNP} Cornean H.D., Duclos P., Nenciu G., and Purice R.:
Adiabatically switched-on electrical bias and the Landauer-B\"uttiker formula.  
J. Math. Phys. {\bf 49}, 102106 (2008).

\bibitem[CDP]{CDP} Cornean H.D., Duclos P., and Purice R.: Adiabatic Non-Equilibrium Steady States in the Partition Free Approach.  
Ann. Henri Poincar\'e {\bf 13}(4), 827-856 (2012).

\bibitem[CGZ]{CGZ} Cornean, H.D., Gianesello, C., and Zagrebnov, V.:
A partition-free approach to transient and steady-state charge currents.
J. Phys. A: Math. Theor. {\bf 43}, 474011 (2010). 

\bibitem[CJM]{CJM} Cornean, H.D., Jensen, A., and Moldoveanu, V.:
A rigorous proof of the Landauer-B\"uttiker formula.
J. Math. Phys. {\bf 46}, 042106, (2005).

\bibitem[CM]{CM} Cornean, H.D., and Moldoveanu, V.: 
On the cotunneling regime of interacting quantum dots. 
J. Phys. A: Math. Theor. {\bf 44}, 305002, (2011). 

\bibitem[CNZ]{CNZ} Cornean H., Neidhardt H. and Zagrebnov V.:
Time-dependent coupling does not change the steady state.
Ann. Henri Poincar\'e {\bf 10}, 61, (2009).

\bibitem[Da1]{Da1}Davies, E.B.:
\newblock Markovian master equations.
\newblock Commun. Math. Phys. {\bf 39}, 91--110 (1974).

\bibitem[Da2]{Da2}Davies, E.B.:
\newblock Markovian master equations. II.
\newblock Math. Ann. {\bf 219}, 147--158 (1976).

\bibitem[Da3]{Da3}Davies, E.B.:
\newblock Markovian master equations. III.
\newblock Ann. Inst. H. Poincar\'e, section B, {\bf 11}, 265--273 (1975).

\bibitem[DeGe]{DeGe} Derezi\'nski, J., and G\'erard, C.:
{\sl Mathematics of Quantization and Quantum Fields.}
Cambridge University Press, Cambridge, UK, 2013.

\bibitem[DFG]{DFG} Dirren, S.: ETH diploma thesis winter 1998/99, chapter 5 (written under
the supervision of J. Fr\"ohlich and G.M. Graf).

\bibitem[DM]{DM} de Roeck, W. and Maes, C.:
Steady state fluctuations of the dissipated heat for a quantum stochastic model.
Rev. Math. Phys. {\bf 18}, 619--654 (2006).

\bibitem[DRM]{DRM} Derezi\'nski, J., de Roeck, W., and Maes, C.:
Fluctuations of quantum currents and unravelings of master equations.
J. Stat. Phys. {\bf 131}, 341--356 (2008).

\bibitem[EHM]{EHM} Esposito M., Harbola U. and Mukamel S.: Nonequilibrium fluctuations, 
fluctuation theorems, 
and counting statistics in quantum systems. Rev. Mod. Phys. {\bf 81}, 1665 
(2009).

\bibitem[Ev]{Ev} Evans, D.E.: Scattering in the CAR algebra.
Commun. Math. Phys. {\bf 48}, 23--30 (1976).


\bibitem[FMSU]{FMSU}  Fr\"ohlich, J., Merkli, M., Schwarz, S.,  and Ueltschi, D.:
Statistical  mechanics of thermodynamic processes. In {\sl A Garden of Quanta
 (Essays in Honor of Hiroshi  Ezawa).} J. Arafune et al. (eds.). World Scientific,
London, Singapore, Hong Kong 2003.

\bibitem[FMU]{FMU} Fr\"ohlich, J., Merkli, M., and Ueltschi, D.: 
Dissipative transport: thermal contacts and tunneling junctions.
Ann. Henri Poincar\'e {\bf 4}, 897 (2004).

\bibitem[FNBSJ]{FNBSJ}
Flindt C., Novotny T., Braggio A., Sassetti M. and Jauho A-P.:                                     
 Counting Statistics of Non-Markovian Quantum Stochastic Processes. Phys. 
Rev. Lett. {\bf 100}, 150601 (2008).

\bibitem[FNBJ]{FNBJ}
Flindt C., Novotny T., Braggio A., and Jauho A-P.: Counting statistics of 
transport through Coulomb 
blockade nanostructures: High-order cumulants and non-Markovian effects. 
Phys. Rev. B {\bf 82}, 155407 (2010).


\bibitem[H]{H} Hewson, A. C.: {\sl The Kondo Problem to Heavy Fermions.}
Cambridge University Press, Cambridge, 1993.

\bibitem[He1]{He1} Hepp, K.: 
Rigorous results on the s--d model of the Kondo effect.
Solid State Communications {\bf 8}, 2087--2090 (1970).

\bibitem[He2]{He2} Hepp, K.: 
Results and problems in irreversible statistical mechanics
of open systems. In {\sl  International Symposium on Mathematical Problems in Theoretical Physics,
January 23--29, 1975, Kyoto University, Kyoto, Japan.} H. Araki editor. Lecture Notes in Physics
{\bf 39}, Springer, Berlin, 1975.

\bibitem[Im]{Im} Imry, Y.: {\sl Introduction to Mesoscopic Physics.}
Oxford University Press, Oxford, 1997.

\bibitem[JK]{JK} Jensen, A., and Kato T.: 
Spectral properties of Schr\"odinger operators and time-decay of the wave functions.
Duke Math. J. {\bf 46}, 583 (1979).

\bibitem[JOP1]{JOP1} Jak\v si\'c, V., Ogata, Y., and Pillet, C.-A.:
The Green-Kubo formula and the Onsager reciprocity relations in quantum statistical mechanics. 
Commun. Math. Phys. {\bf 265}, 721--738 (2006).

\bibitem[JOP2]{JOP2} Jak\v si\'c, V., Ogata, Y., and Pillet, C.-A.:
Linear response theory for thermally driven quantum open systems.
J. Stat. Phys. {\bf 123}, 547--569 (2006).

\bibitem[JOP3]{JOP3} Jak\v si\'c, V., Ogata, Y., and Pillet, C.-A.:
The Green-Kubo formula for locally interacting fermionic open systems.
Ann. Henri Poincar\'e {\bf 8}, 1013--1036 (2007).

\bibitem[JOPP]{JOPP} Jak\v si\'c, V., Ogata, Y., Pautrat, Y., and Pillet, C.-A.:
Entropic fluctuations in quantum statistical mechanics -- an introduction. 
In {\sl Quantum Theory from Small to Large Scales.} 
J. Fr\"ohlich, M. Salmhofer, W. de Roeck, V. Mastropietro and L.F. Cugliandolo editors. 
Oxford University Press, Oxford, 2012.

\bibitem[JOPS]{JOPS} Jak\v si\'c, V., Ogata, Y., Pillet, C.-A., and Seiringer, R.:
Quantum hypothesis testing and non-equilibrium statistical mechanics.
Rev. Math. Phys. {\bf 24}, 1230002 (2012).

\bibitem[JP1]{JP1} Jak\v si\'c, V., and Pillet, C.-A.:
Non-equilibrium steady states of finite quantum systems coupled to thermal reservoirs.
Commun. Math. Phys. {\bf 226}, 131--162 (2002).

\bibitem[JP2]{JP2} Jak\v si\'c, V., and Pillet, C.-A.:
Mathematical theory of non-equilibrium quantum statistical mechanics.
J. Stat. Phys. {\bf 108}, 787--829 (2002).

\bibitem[JP3]{JP3} Jak\v si\'c, V., and Pillet, C.-A.:
On entropy production in quantum statistical mechanics.
Commun. Math. Phys. {\bf 217}, 285--293 (2001).

\bibitem[JP4]{JP4} Jak\v si\'c, V., and Pillet, C.-A.:
A note on the entropy production formula.
Contemp. Math. {\bf 327}, 175--180 (2003).

\bibitem[JP5]{JP5} Jak\v si\'c, V., and Pillet, C.-A.:
On the strict positivity of entropy production.
In {\sl Adventures in Mathematical Physics - Transport and Spectral Problems in 
Quantum Mechanics: A Conference in Honor of Jean-Michel Combes}. 
F.~Germinet and  P.D.~Hislop editors.
Contemp. Math. {\bf 447}, 153--163 (2007).

\bibitem[JPP]{JPP} Jak\v si\'c, V., Pautrat, Y., and Pillet, C.-A.:
Central limit theorem for locally interacting Fermi gas.
Commun. Math. Phys. {\bf 285}, 175--217 (2009).

\bibitem[JWM]{JWM} Jauho, A.-P., Wingreen, N.S., and Meir Y.:
Time-dependent transport in interacting and noninteracting resonant-tunneling systems.
Phys. Rev. B {\bf 50}, 5528 (1994).

\bibitem[KAO]{KAO} Kashcheyevs, V, Aharony, A., and Entin-Wohlman, O.:
Applicability of the equations-of-motion technique for quantum dots.
Phys. Rev. B {\bf 73}, 125338 (2006). 

\bibitem[Ke]{Ke} Keldysh, L.V.: 
Diagram technique for nonequilibrium processes.
Zh. Eksp. Teor. Fiz. {\bf 47}, 1515 (1964). English translation in 
Sov. Phys. JETP {\bf 20}, 1018 (1965).

\bibitem[KSKVG]{KSKVG} Kurth, S., Stefanucci, G., Khosravi, E., Verdozzi, C., and Gross, E.K.U.:
Dynamical Coulomb Blockade and the Derivative Discontinuity of Time-Dependent Density 
Functional Theory.
Phys. Rev. Lett. {\bf 104}, 236801 (2010).

\bibitem[LL]{LL} Levitov, L.S., and Lesovik, G.B.:
Charge distribution in quantum shot noise. 
JETP Lett. {\bf 58}, 230--235 (1993).

\bibitem[LLL]{LLL}
 Levitov L. S., Lee H., and Lesovik G. B.: Electron counting statistics 
and coherent states 
of electric current. J. Math. Phys. {\bf 37}, 4845 (1996).

\bibitem[LS]{LS} Lebowitz, J.L., and Spohn, H.: 
Irreversible thermodynamics for quantum systems weakly coupled to thermal reservoirs.
Adv. Chem. Phys. {\bf 38}, 109--142 (1978).

\bibitem[MCP]{MCP} Moldoveanu, V., Cornean, H.D., and Pillet C.-A.:  
Non-equilibrium steady-states for interacting open systems: exact results. 
Phys. Rev. B. {\bf 84}, 075464, (2011). 

\bibitem[MMS]{MMS} Merkli, M., M\"uck, M., and Sigal, I.M.:
Theory of non-equilibrium stationary states as a theory of resonances.
Ann. Henri Poincar\'e {\bf 8}, 1539--1593 (2007).

\bibitem[MSSL]{MSSL} Myohanen, P., Stan, A., Stefanucci, G., and van Leeuwen, R.:
Kadanoff-Baym approach to quantum transport through interacting nanoscale systems: 
From the transient to the steady-state regime.
Phys. Rev. B {\bf 80}, 115107 (2009).

\bibitem[MW]{MW} Meir, Y., and Wingreen, N.S.: 
Landauer formula for the current through an interacting electron region. 
Phys. Rev. Lett. {\bf 68}, 2512 (1992).

\bibitem[Ne]{Ne} Nenciu, G.: 
Independent electrons model for open quantum systems: Landauer-B\"uttiker formula and strict
positivity of the entropy production.
J. Math. Phys. {\bf 48}, 033302 (2007).

\bibitem[NT]{NT} Narnhofer, H., and Thirring, W.:
Adiabatic theorem in quantum statistical mechanics.
Phys. Rev. A {\bf 26}, 3646 (1982).

\bibitem[P]{P} Pearson, D.B.:
{\sl Quantum Scattering and Spectral Theory.}
Academic Press, London, 1988.

\bibitem[Pi]{Pi} Pillet, C.-A.:
Quantum dynamical systems. In {\sl Open Quantum Systems I.}
S. Attal, A. Joye and C.-A. Pillet editors.
Lecture Notes in Mathematics, volume 1880, Springer Verlag, Berlin, 2006.

\bibitem[PFVA]{PFVA} Puig, M. von Friesen, Verdozzi, V., and Almbladh, C.-O.: 
Kadanoff-Baym dynamics of Hubbard clusters: Performance of many-body schemes, 
correlation-induced damping and multiple steady and quasi-steady states. 
Phys. Rev. B {\bf 82}, 155108 (2010).

\bibitem[PG]{PG} Pustilnik, M., and Glazman, L.: 
Kondo effect in quantum dots.
J. Phys. Condens. Matter {\bf 16} R513, (2004).

\bibitem[RS2]{RS2} Reed, M., and Simon, B.:
{\sl Methods of Modern Mathematical Physics. II: Fourier Analysis, Self-Adjointness.}
Academic Press, New York, 1975.

\bibitem[RS3]{RS3} Reed, M., and Simon, B.:
{\sl Methods of Modern Mathematical Physics. III: Scattering Theory.}
Academic Press, New York, 1979.

\bibitem[Ro]{Ro} Robinson, D.W.: 
Return to equilibrium. 
Commun. Math. Phys. {\bf 31}, 171--189 (1973).

\bibitem[Ru1]{Ru1} Ruelle, D.: 
Natural nonequilibrium states in quantum statistical mechanics. 
J. Stat. Phys.{\bf 98}, 57--75 (2000).

\bibitem[Ru2]{Ru2} Ruelle, D.: 
Entropy production in quantum spin systems.
Commun. Math. Phys. {\bf 224}, 3--16 (2001).

\bibitem[Sp]{Sp} Spohn, H.:
An algebraic condition for the approach to equilibrium of an open N-level system.
Lett. Math. Phys. {\bf 2}, 33--38 (1977).

\bibitem[TR]{TR} Thygesen, K.S., and Rubio, A.:
Conserving GW scheme for nonequilibrium quantum transport in molecular contacts.
Phys. Rev. B {\bf 77}, 115333 (2008). 

\end{thebibliography}
\end{document}